\documentclass[aps,prl,twocolumn,amsmath,amssymb,nofootinbib,preprintnumbers]{revtex4}

\usepackage{amsmath,latexsym,amssymb,amsfonts}
\usepackage[dvips]{graphicx,color}
\usepackage{bm}
\usepackage{gensymb}
\usepackage{ulem}
\usepackage[dvipsnames]{xcolor}

\begin{document}

\title{\textbf{AnaBHEL (Analog Black Hole Evaporation via Lasers) Experiment: \\
Concept, Design, and Status}}

\author{
[AnaBHEL Collaboration] \\
Pisin Chen$^{a,b,c}$\footnote{{\tt pisinchen{@}phys.ntu.edu.tw}}, 
Gerard Mourou$^{d}$, 
Marc Besancon$^{e}$,
Yuji Fukuda$^{f}$, 
Jean-Francois Glicenstein$^{e}$,
Jiwoo Nam$^{a,b,c}$, 
Ching-En Lin$^{a,b}$, 
Kuan-Nan Lin$^{a,b}$, 
Shu-Xiao Liu$^{a}$,
Yung-Kun Liu$^{a,b}$, 
Masaki Kando$^{f}$, 
Kotaro Kondo$^{f}$, 
Stathes Paganis$^{a,b}$, 
Alexander Pirozhkov$^{f}$, 
Hideaki Takabe$^{a}$, 
Boris Tuchming$^{e}$,
Wei-Po Wang$^b$,
Naoki Watamura$^{a}$, 
Jonathan Wheeler$^{d}$, 
Hsin-Yeh Wu$^{a,b}$ 
}

\affiliation{
$^{a}$Leung Center for Cosmology and Particle Astrophysics, National Taiwan University, Taipei 10617, Taiwan\\
$^{b}$Department of Physics, National Taiwan University, Taipei 10617, Taiwan\\
$^{c}$Graduate Institute of Astrophysics, National Taiwan University, Taipei 10617, Taiwan\\
$^{d}$IZEST, Ecole Polytechnique, Palaiseau, France\\
$^{e}$Irfu, CEA, Université Paris-Saclay, F-91191 Gif sur Yvette, France\\
$^{f}$Kansai Photon Science Institute, National Institutes for Quantum Science and Technology, 8-1-7 Umemidai, Kizugawa, Kyoto 619-0215, Japan
\\
}

\begin{abstract}
Accelerating relativistic mirror has long been recognized as a viable setting where the physics mimics that of black hole Hawking radiation. In 2017, Chen and Mourou proposed a novel method to realize such a system by traversing an ultra-intense laser through a plasma target with a decreasing density. An international AnaBHEL (Analog Black Hole Evaporation via Lasers) Collaboration has been formed with the objectives of observing the analog Hawking radiation and shedding light on the information loss paradox. To reach these goals, we plan to first verify the dynamics of the flying plasma mirror and to characterize the correspondence between the plasma density gradient and the trajectory of the accelerating plasma mirror. We will then attempt to detect the analog Hawking radiation photons and measure the entanglement between the Hawking photons and their ``partner particles". In this paper, we describe our vision and strategy of AnaBHEL using the Apollon laser as a reference, and we report on the progress of our R\&D of the key components in this experiment, including the supersonic gas jet with a graded density profile, and the superconducting nanowire single-photon Hawking detector. In parallel to these hardware efforts, we performed computer simulations to estimate the potential backgrounds, and derive analytic expressions for modifications to the blackbody spectrum of Hawking radiation for a perfectly reflecting, point mirror, due to the semit-ransparency and finite-size effects specific to flying plasma mirrors. Based on this more realistic radiation spectrum, we estimate the Hawking photon yield to guide the design of the AnaBHEL experiment, which appears to be achievable.

\end{abstract}

\maketitle

\section{Introduction}

The question of whether Hawking evaporation \cite{Hawking:1974sw} violates unitarity, and therefore results in the loss of information \cite{Hawking:1976}, remains unresolved since Hawking's seminal discovery. Proposed solutions range from black hole complementarity [3], firewalls \cite{Almheiri:2012a, Almheiri:2012b} (see, for example, \cite{Mathur:2009, Chen:2015gux}, for a recent review and \cite{Chen:2015gux, Buosso:2017, Giddings:2019} for a counter argument), soft hairs \cite{Hawking:2015}, black hole remnants \cite{Chen:2014jwq}, islands \cite{Almheiri:2019psf, Almheiri:2019hni}, and replica wormholes \cite{Penington:2019kki, Almheiri:2019qdq}, to instanton tunneling between multiple histories of Euclidean path integrals \cite{Chen:2021jzx}.  So far the investigations remain mostly theoretical since it is almost impossible to settle this paradox through direct astrophysical observations, as typical stellar-size black holes are cold and young, yet the solution to the paradox depends crucially on the end-stage of the black hole evaporation.

There have been proposals for laboratory investigations of the Hawking effect, including sound waves in moving fluids \cite{Unruh:1981}, electromagnetic waveguides \cite{Schuthold:2005}, traveling index of refraction in media \cite{Yablonovitch:1989}, ultra-short laser pulse filament \cite{Belgiorno:2010}, Bose-Einstein condensates \cite{deNova:2019}, and electrons accelerated by intense lasers \cite{Chen-Tajima}. In particular, Ref.\cite{deNova:2019} reported on the observation of a thermal spectrum of Hawking radiation in the analog system and its entanglement. However, most of these are limited to verifying the thermal nature of the Hawking radiation. 

It has long been recognized that accelerating mirrors can mimic black holes and emit Hawking-like thermal radiation \cite{Fulling:1976}. In 2017, Chen and Mourou proposed a scheme to physically realize a relativistic mirror by using a state-of-the-art high intensity laser to impinge a plasma target with a decreasing density \cite{Chen:2017,Chen:2020}. The proposal is unique in that it does not rely on certain fluid to mimic the curved spacetime around a black hole, but rather a more direct quantum field theoretical analogy between the spacetime geometry defined by a black hole and a flying mirror. 

Based on this concept, an international AnaBHEL Collaboration has been formed to carry out the Chen-Mourou scheme, which is the only experimental proposal of its kind in the world. 
Our ultimate scientific objectives are to detect analog Hawking radiation for the first time in history and through the measurement of the quantum entanglement between the Hawking particles and their vacuum fluctuating pair partner particles, to shed some light on the unresolved information loss paradox. From this perspective, the AnaBHEL experiment may be regarded as a {\it flying} EPR (Einstein-Podolsky-Rosen) experiment \cite{EPR}.

The concept of flying plasma mirror was proposed by Bulanov et al. \cite{Bulanov:2003, Naumova:2004, Bulanov:2013}. It provides an alternative approach to the free electron laser (FEL) in generating high frequency coherent radiation. The flying plasma mirror approach provides a great prospect for future applications. A series of proof-of-principle experiments led by Kando at KPSI, Japan \cite{Pirozhkov:2007, Kando:2007, Kando:2009} has validated the concept. However, the mirror reflectivity (as a function of frequency) as well as other physical properties such as the reflection angular distribution, etc., have not been characterized in those two experiments.

In this paper, we first review the physics of flying mirror as an analog black hole. We then reveal the concept of accelerating relativistic plasma mirrors as analog black holes, with the attention paid to the aspects pertinent to the investigation of Hawking radiation and the information loss paradox, including the laser-plasma dynamics that gives rise to the acceleration of the plasma mirror, the reflectivity and the frequency-shift of the reflected spectrum, and corrections due to the finite-size and the semitransparency effects of a realistic plasma mirror to  the blackbody spectrum of the analog Hawking radiation based on an idealized, perfectly reflecting, point mirror. We then report on the progress of our R\&D of the key components in the AnaBHEL experiment, including that of the supersonic gas jet and the superconducting nanowire single-photon Hawking detector. We conclude by projecting our experimental outlook.

\section{Flying Mirror as Analog Black Hole}

Figure 1 depicts the analogy between the Hawking radiation of a real BH (left) and that of an accelerating mirror (right). 
That accelerating mirrors can also address the information loss paradox was first suggested by Wilczek \cite{Wilczek:1992}. As is well-known, the notion of black hole information loss is closely associated with quantum entanglement. In order to preserve the ``black hole unitarity", Wilczek argued, based on the moving mirror model, that in vacuum fluctuations the {\it partner modes} of the Hawking particles would be trapped by the horizon until the end of the evaporation, where they would be released and the initial pure state of a black hole would be recovered with essentially zero cost of energy. More recently, Hotta et al. \cite{Hotta:2015} argued that the released partner modes are simply indistinguishable from the zero-point vacuum fluctuations. On the other hand, there is also the notion that these partner modes would be released in a burst of energy, for example, in the Bardeen model \cite{Bardeen:2014} (See Fig. 2). 

One common {\it drawback} of all analog black hole concepts is that, being setting up in a laboratory with flat spacetime and therefore the standard quantum field theory is known to be valid, it is inevitable that any physical process, including analog black hole systems, must preserve the unitarity. 
Therefore none of the proposed analog black holes can in principle {\it prove} the loss of information even if that is indeed so. The real issue is therefore not so much about whether the unitarity is preserved, but more about {\it how} it is preserved. That is, it is even more important to determine how the black hole information is retrieved. Does it follow the Page curve \cite{Page:1993}, or a modified Page curve where the Page time is significantly shifted towards the late time \cite{Chen:2021jzx}, or some alternative scenarios \cite{Hotta-Sugita:2015}? The measurement of the entanglement between the Hawking particles and the partner particles as well as the evolution of the entanglement entropy \cite{Chen-Yeom:2017}, should help to shed much light on the black hole information loss paradox. As has been pointed out by Chen and Yeom  \cite{Chen-Yeom:2017}, different scenarios of black hole evolution can be tested by different mirror trajectories \cite{Good:2020}.

\begin{figure}
\begin{center}
\includegraphics[scale=0.80]{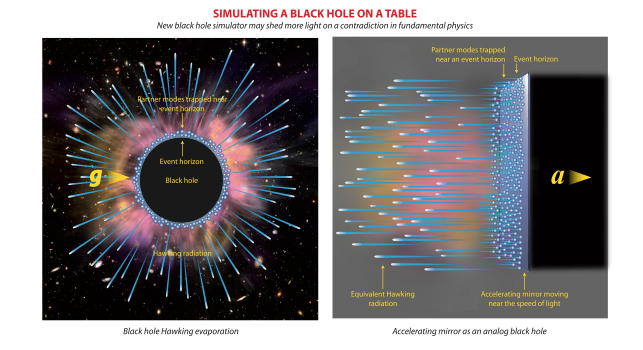}
\caption{\label{fig:Picture1.png}The analogy between the Hawking radiation from a true BH (left) and that from an accelerating mirror (right). One may intuitively appreciate their analogy based on Einstein’s Equivalence Principle.}
\end{center}
\end{figure}

\begin{figure}
\begin{center}
\includegraphics[scale=0.80]{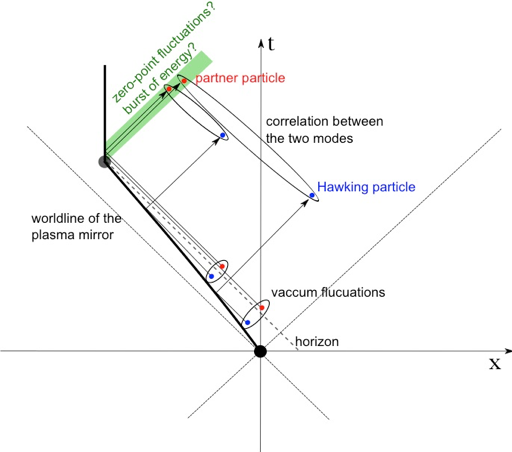}
\caption{\label{fig:Picture0.png}
(Reproduced from Ref.\cite{Chen:2017}.) The worldline of an accelerating relativistic plasma mirror and its relation with vacuum fluctuations around the horizon. In particular, the entanglement between the Hawking particles (blue) emitted at early times and their partner particles (red) collected at late times is illustrated. The green strip represents either a burst of energy or zero-point 
fluctuations emitted when the acceleration stops abruptly.}
\end{center}
\end{figure}

\section{Accelerating Plasma Mirror via Density Gradient}

As is well known, plasma wakefields \cite{Tajima:1979, Chen:1985} in the nonlinear regime of laser-plasma interaction will blow out all the intervening plasma electrons, leaving an “ion bubble” trailing behind the driving laser pulse or electron beam. Eventually the expelled electrons will rush back and pile up with a singular density distribution. S. Bulanov et al. \cite{Bulanov:2003, Naumova:2004, Bulanov:2013} suggested that such a highly nonlinear plasma wake could serve as a relativistically flying mirror where an optical frequency light, upon reflecting from the flying plasma mirror, would instantly blueshift to an X-ray. To apply this flying plasma mirror concept to the investigation of black hole Hawking evaporation, one must make the plasma mirror accelerate.

In this regard, one important issue is the correspondence between the plasma density gradient and the mirror spacetime trajectory. In order to mimic the physics of Hawking evaporation, the plasma mirror must undergo a non-trivial acceleration that gives rise to a spacetime trajectory that is black hole physics meaningful. Such black hole relevant trajectories have been well studied theoretically in the past 40 years with a wealth of literature available. This, as proposed \cite{Chen:2017,Chen:2020}, can be realized by preparing the plasma target with a prescribed density gradient (See Fig. 3 for a schematic drawing of the concept).

\begin{figure}
\begin{center}
\includegraphics[scale=0.80]{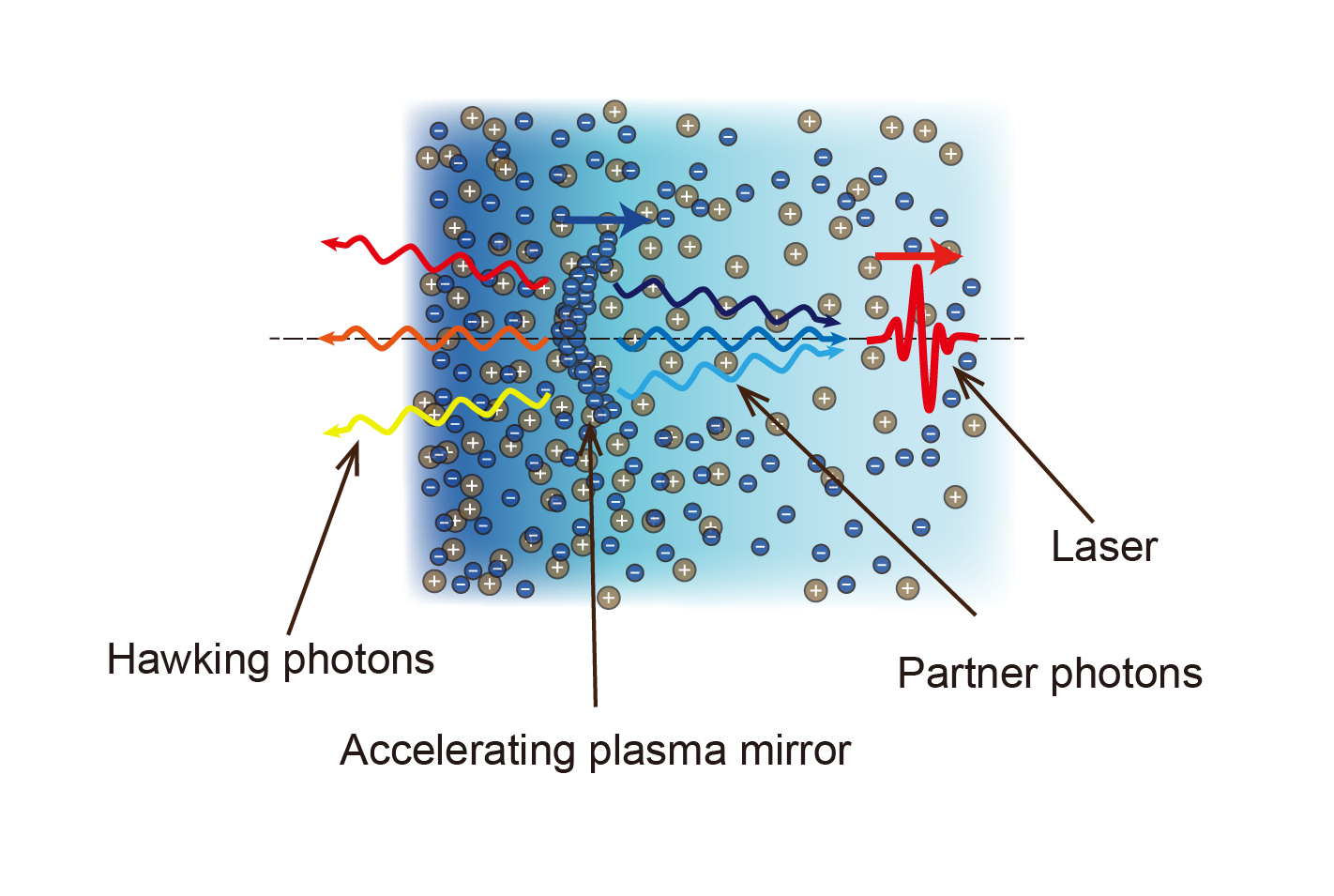}
\caption{\label{fig:Picture8.png}
Schematic drawing of the concept of accelerating plasma mirror driven by an intense laser that traverses a plasma with a decreasing density. Due to the variation of the laser intensity in the transverse dimension, which is typically in Gaussian distribution, the flying plasma mirror induced by the laser is concave in the forward direction.}
\end{center}
\end{figure}

\subsection{Mirror Trajectory and Plasma Density Correspondence}
In Ref.\cite{Chen:2020}, the acceleration and thus the trajectory of a plasma mirror as a function of the local plasma density and gradient was derived. In general, 
\begin{eqnarray}
\label{eq:plasma_mirror_accleration_general}
\frac{\ddot{x}_{_M}}{c^2}&=&-\frac{1-(1/2)\omega_p^2/\omega_0^2}{[1+(3\pi/2) c\omega_p'/\omega_p^2]^3} \times \\  \nonumber
&&\Big\{\frac{\omega_p^2}{\omega_0^2}\frac{\omega_p'}{\omega_p}
+\frac{3\pi c}{2}\Big[\frac{\omega_p''}{\omega_p^2}-2\frac{\omega_p'^2}{\omega_p^3}\Big]\Big\},
\end{eqnarray}
where ${x}_{_M}$ is the position of the plasma mirror, $\ddot{x}_{_M}$ its second time derivative, $\omega_0$ the laser frequency,  $\omega_p(x)=c\sqrt{4\pi r_e n_p(x)}$ the local plasma frequency, $r_e=e^2/m_ec^2$ the classical electron radius, and $\omega_p'\equiv \partial\omega_p(x)/\partial x$. Our desire is to achieve as high an acceleration as possible. To accomplish that, one should design the system in such a way that the denominator of Eq.(1) is minimized. 

A simple but well motivated plasma density profile is the one that corresponds to the exponential trajectory investigated by Davies and Fulling \cite{Fulling:1976,Davies:1977}, which is of special geometrical interest because it corresponds to a well-defined horizon \cite{Birrell:1982ix}. Inspired by that, we consider the following plasma density variation along the direction of the laser propagation inside the plasma target with thickness $L$:
\begin{eqnarray}
n_p(x)=n_{p0}(a+be^{x/D})^2, \quad\quad -L\leq x  \leq 0,
\end{eqnarray}
where $n_{p0}(a+b)^2$ is the plasma density at $x=0$, $D$ is the characteristic length of density variation. Accordingly, the plasma frequency varies as 
\begin{eqnarray}
\omega_p(x)=\omega_{p0}(a+be^{x/D}), \quad\quad -L\leq x  \leq 0,
\end{eqnarray}
where $\omega_{p0}=c\sqrt{4\pi r_e n_{p0}}$. In our conception, the time derivatives of the plasma frequency are induced through the spatial variation of the plasma density via the relation $\omega_p(x)=c\sqrt{4\pi r_e n_p(x)}$. Thus
\begin{eqnarray}
\omega'_p(x)&=&\frac{b}{D}e^{x/D}\omega_{p0}, \\ 
\omega''_p(x)&=&\frac{b}{D^2}e^{x/D}\omega_{p0}.
\end{eqnarray}
Inserting these into Eq.(\ref{eq:plasma_mirror_accleration_general}), we then have, for the constant-plus-exponential-squared distribution of Eq.(2), 
\begin{eqnarray}
\frac{\ddot{x}_{_M}}{c^2}&=&-\frac{1-(\omega_{p0}^2/2\omega_0^2)(a+be^{x/D})^2}
{[1+(3b/4)(\lambda_{p0}/D)e^{x/D}/(a+be^{x/D})^2]^3}  \cr    
&&\times \frac{\lambda_{p0}}{D}\frac{be^{x/D}}{(a+be^{x/D})} \Big\{\frac{\omega_p^2}{\omega_0^2}\frac{1}{\lambda_{p0}} \cr  
&&+ \frac{3}{4D}\Big[\frac{1}{a+be^{x/D}}-\frac{2be^{x/D}}{(a+be^{x/D})^2}\Big]\Big\}. 
\end{eqnarray}

PIC simulations of the laser-plasma interaction were performed based on the above plasma density profile \cite{Liu:2020}. The acceleration of the plasma mirror agrees well with the formula (See Fig. 4).

\begin{figure}
\begin{center}
\includegraphics[scale=0.80]{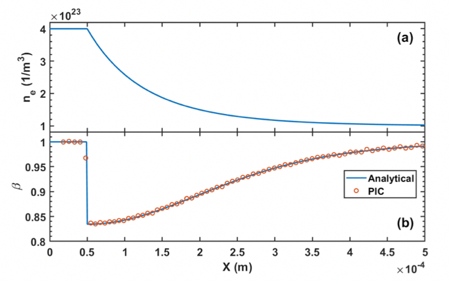}
\caption{\label{fig:Picture2.png}PIC simulation of plasma mirror acceleration \cite{Liu:2020}. (a) A plasma target with a constant-plus-exponential density gradient, $n_p (x)=n_{p0}(1+e^{x/D})^2, n_p (x=0)=1.0\times 10^{17}$${\rm cm}^{-3}$. (b) Comparison of the plasma mirror speed, $\beta=\dot{x}_M/c$, between the analytic formula (solid blue curve) and the PIC simulations (orange circles). The two agree extremely well. Note that the convention of the laser propagation direction in this PIC simulation is from left to right, which is opposite to that in the typical theoretical treatment of flying mirrors as analog black holes.}
\end{center}
\end{figure}

\section{\label{sec:max}Analog Hawking Temperature}

There exists a wealth of literature on the vacuum fluctuating modes of quantum fields, their reflections from a flying mirror, and the analog ``Hawking temperature" of such a flying mirror as an analog black hole \cite{Birrell:1982ix}. In general, such analog Hawking temperature depends on the actual mirror trajectory. According to Ref.[25],
\begin{eqnarray}
\int_{0}^{t}\bar{c} dt=
\int^{0}_{x_{_M}}dx \Big[1+\frac{3b\lambda_{p0}}{4D}\frac{e^{x/D}}{(a+be^{x/D})^2}\Big],
\end{eqnarray} 
where $\bar{c}=\eta c=(1-\omega_{p}^2/2\omega_0^2)c$ is the speed of light in the plasma medium, which is position dependent.
In our conception \cite{Chen:2017}, the plasma target thickness is supposed to be much larger than the characteristic scale of the density variation, i.e., $L\gg D$. In this situation one is safe to extend the integration to $x_{_M}\to -\infty$ (and $t\to \infty$). Taking this approximation, we find 
\begin{eqnarray}
x_{_M}(t)=-\eta_act-Ae^{-\eta_act/D}+A,  \quad  t \to \infty,
 \end{eqnarray}
where $\eta_a=1-a^2\omega_{p0}^2/2\omega_0^2$ and
$A=\eta_aD[ab(\omega_{p0}^2/\omega_0^2)-(3b/4a^2)(\lambda_{p0}/D)]$.
This is identical to the Davies-Fulling trajectory, i.e., Eq(4.51) of Ref.\cite{Birrell:1982ix},
\begin{eqnarray}
z(t) \to -t-Ae^{-2\kappa t}+B, \quad\quad t \to \infty,
 \end{eqnarray}
where $A,B,\kappa$ are positive constants and $c\equiv1$. 

Transcribing the $x_{_M}(t)$ coordinates to the $(u,v)$ coordinates, where $u=\eta_act-x_{_M}(t)$ and $v=\eta_act+x_{_M}(t)$, we see that only null rays with $v<A$ can be reflected. All rays with $v>A$ will pass undisturbed. The ray $v=A$ therefore acts as an effective horizon \cite{Birrell:1982ix}. Following the standard recipe \cite{Birrell:1982ix}, we obtain the Wightman function as 
\begin{eqnarray}
D^+(u,v;u',v')=&-&\frac{1}{4\pi}\ln\big[2Ae^{2\eta_ac(t+t')/2D} \cr
&\times& \sinh(\eta_ac\Delta t/2D)\big],
\end{eqnarray}
where $\Delta t=t-t'=\Delta u/2\eta_ac$ in the $t\to \infty$ limit. The constant factors in the argument of the log function in the above equation do not contribute to the nontrivial part of the physics. Note that in our notation $t$ is the time when the ray hits the mirror. Let us denote the observation time and position by $T$ and $X$. Then $u=\eta_acT-X=\eta_act-x_{_M}$. For large $t$, $u=\eta_acT-X=2\eta_act-A$. This leads to $\Delta u=2\eta_ac\Delta t=\eta_ac\Delta T$ for a static mirror at $X={\rm const.}$ Integrating over $T$ and $T'$, we then have, in the asymptotic limit of $t, t' \to \infty$, 
\begin{eqnarray}
D^+(u,v;u',v')=-\frac{1}{4\pi}\ln\big[\sinh(\eta_ac\Delta t/2D)\big].
\end{eqnarray} 
This leads to the response function (of the particle detector) per unit time with the form
\begin{eqnarray}
\mathcal{F}(E)/{\rm unit \ time}=\frac{1}{E}\frac{1}{(e^{E/k_{_B}T_{H}}-1)},
\end{eqnarray}
where the analog Hawking temperature of the mirror measured by a stationary particle detector is
\begin{eqnarray}
k_{_B}T_{_H}=\frac{\hbar c}{4\pi}\frac{\eta_a}{D}.
\end{eqnarray}
Here $k_{_B}$ is the Boltzmann constant. It is interesting to note that the analog Hawking temperature associated with our constant-plus-exponential-squared density profile depends strongly on the characteristic length $D$ and only weakly on the plasma density (through $\eta_a$). This points to the possibility of employing gaseous instead of solid plasma targets, which would greatly simplify our proposed experiment.

\section{Conceptual Design}
The original experimental concept proposed by Chen and Mourou \cite{Chen:2017} invoked a two-plasma-target approach, where the first plasma target converts an optical laser into an X-ray pulse through the flying plasma mirror mechanism. The converted X-ray pulse then impinges on a nano-thin-film that is fabricated with a graded density in different layers. 
This design has the advantage of having the solid state density to provide a higher plasma frequency, which is proportional to the square-root of the plasma density, and therefore a higher density gradient for maximizing Hawking temperature. On the other hand, the drawback of this concept are multiple. First, the typical conversion efficiency of flying plasma mirrors is $\sim 10^{-5}$, rendering it difficult for the converted X-ray pulse to remain in the nonlinear regime. Second, the solid plasma target would induce extra backgrounds, which is linearly proportional to the target density. 

In 2020, Chen and Mourou proposed a second design concept \cite{Chen:2020}, where the conversion of optical laser to X-ray is no longer needed and thus the first plasma target was removed, and the nano-thin-film solid plasma target was replaced by a supersonic gas jet. This largely simplifies the design and the technical challenges. Figure 5 shows a schematic conceptual design of the single-target, optical laser approach. The key components now reduce to a supersonic gas jet with a graded density profile and a superconducting nanowire single-photon Hawking detector, the R\&D progress of which will be described in later sections. 

In our design of the AnaBHEL experiment, we assume the driving laser has the frequency $\omega_0=3.5 \times 10^{15}$ ${\rm sec}^{-1}$ and the wavelength $\lambda_p=540$ ${\rm nm}$. For the plasma target, we set $a=b=1$ in Eq.(2) so that $n_p(x)=n_{p0}(1+e^{x/D})^2$, and we assume $n_{p}(x=0)=1.0 \times 10^{17}$ ${\rm cm}^{-3}=4n_{p0}$. The corresponding plasma frequency is $\omega_{p0}=0.9 \times 10^{13}$ ${\rm sec}^{-1}$ and the plasma wavelength $\lambda_{p0}=200$ $\mu{\rm m}$. Next we design the plasma target density profile. Since our formula is not constrained by the adiabatic condition, we are allowed to choose a minute characteristic length $D=0.5$ $\mu{\rm m}$. Then we find
\begin{eqnarray}
k_{_B}T_{_H}\sim 3.1 \times 10^{-2}\ {\rm eV},
\end{eqnarray}
which corresponds to a characteristic Hawking radiation frequency $\omega_{_H} \sim 4.8\times 10^{13}$ ${\rm sec}^{-1} > \omega_{p0}$. So the Hawking photons can propagate through and out of the plasma for detection.

\begin{figure*}[t]
\begin{center}
\includegraphics[scale=0.80]{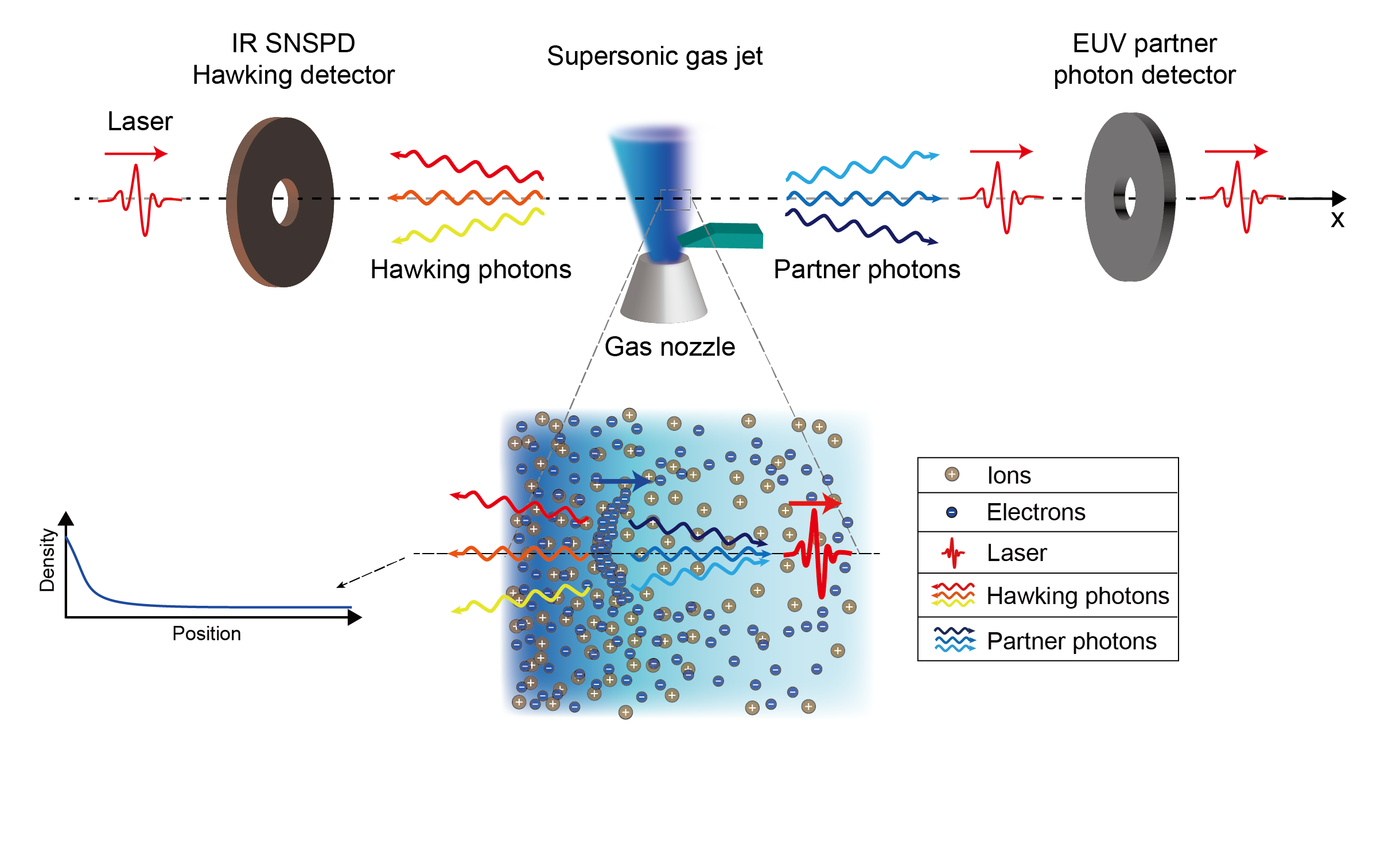}
\caption{\label{fig:Picture6.png}     
A conceptual design of the AnaBHEL experiment. The enlarged figure is a gaseous plasma target with a decreasing density profile where the penetrating optical laser pulse (red) would induce an accelerating flying plasma mirror (blue). Hawking photons would be emitted to the backside of the mirror, which suffer Doppler red-shift and would be in the infrared range. The partner photons, on the other hand, would penetrate the semi-transparent plasma mirror and propagate in the same direction as that of the laser, which do not suffer Doppler red-shift and would be in the EUV range.}
\end{center}
\end{figure*}

\section{Hawking Photon Yield}

Among the proposed models, the physics of flying/moving mirrors is perhaps the one closest to that of real black holes, since in both cases the radiation is originated from vacuum fluctuations. The essence of Hawking radiation lies in the gravitational red-shift of the field modes' phases. Since the key is the phase shift, various analog models or experimental proposals attempt to generate the same phase shift as that of Hawking radiation but now in flat spacetime, i.e., laboratory. Indeed, in the flying mirror model, the gravitational red-shift is mimicked by the Doppler red-shift.

Due to the spherically symmetric nature of typical black hole spacetimes, the spherical coordinate origin is effectively a perfectly reflecting point mirror and the corresponding Hawking radiation is expected to be emitted radially, hence the situation is effectively (1+1)-dimensional and thus most of the flying mirror literature only considers a real perfectly reflecting point mirror in (1+1)-dimensional flat spacetime. Nevertheless, in the laboratory, the spacetime is (1+3)-dimensional. In addition, our proposed relativistic flying mirror generated through laser-plasma interaction has a low reflectivity \cite{Chen:2020} and a finite transverse/longitudinal size, it is therefore necessary to take these practical effects into consideration to estimate the particle production yield.

The standard treatment in the flying mirror model \cite{DeWitt1975,Fulling:1976,Davies:1977,Birrell:1982ix} considers a real scalar field in (1+1)D flat spacetime subjected to a single, relativistic, time-dependent Dirichlet boundary condition in space to represent a relativistic perfectly reflecting point mirror. Since the boundary condition is externally provided, the breakdown of Poincaré invariance leads to the possibility of spontaneous particle creations following quantum field theory.

The generalization of this standard calculation to a flying plasma mirror with a finite reflectivity in $n$-dimensional flat spacetime can be made by starting with the action functional \cite{Barton1995}:
\begin{equation}
\begin{aligned}
    S_{\mu}[\phi]=&-\frac{1}{2}\int_{-\infty}^{\infty}d^nx\;\partial^{\mu}\phi(x)\partial_{\mu}\phi(x)
    \\
    &-\frac{\mu}{2}\int_{-\infty}^{\infty}d^nx\;V(x)\phi^2(x),
\end{aligned}
\end{equation}
where natural units are employed, $\mu=4\pi n_{s}\alpha/m_{e}$ is the coupling constant with dimension of mass, $\alpha=1/137$ is the fine structure constant, $n_{s}$ is the surface density of electrons on the mirror, and
\begin{equation}
    V(x)=\gamma^{-1}(t)H(\mathbf{x}_\perp)f(x-x_{_M}(t)),
\end{equation}
encodes the mirror's trajectory $x_{_M}(t)$, longitudinal/transverse distribution $H/f$, and the Lorentz factor $\gamma$.

Solving the equation of motion for $\phi$ with the in-mode/out-mode boundary conditions in (1+1) dimensions, one finds, assuming the field to be in the in-vacuum state $\left|0;\text{in}\right>$ and the mirror is flying to the negative $x$-direction, the created particles due to the field mode reflected to the mirror's right to have the frequency spectrum \cite{Nicolaevici2001,Nicolaevici2009,Lin2020,Lin2021}:
\begin{equation}
    N=\int_{0}^{\infty}d\omega\int_{0}^{\infty}d\omega'\left|\beta_{\omega\omega'}\right|^2,
\end{equation}
where
\begin{equation}
    \beta_{\omega\omega'}=-\frac{\omega}{2\pi\sqrt{\omega\omega'}}\int_{-\infty}^{\infty}du\;\mathcal{R}_{\omega'}(u)e^{-i\omega'p(u)-i\omega u},
    \label{exact beta 2d}
\end{equation}
and $\omega'/\omega$ is the incident/emitted plane wave mode's frequency, $\mathcal{R}$ is the mirror's reflectivity, $u=t-x_{_M}(t)$, and $p(u)=t+x_{_M}(t)$ is the phase shift/ray-tracing function induced upon reflection off the receding mirror. From Eq.\eqref{exact beta 2d}, one sees that for a given trajectory $x_{_M}$, the spectrum would be different depending on the reflectivity.

A simple model that mimics the formation and evaporation of a Schwarzschild black hole is the collapse of a spherical null shell. In this scenario, the relevant ray-tracing function is
\begin{equation}
    u=p(u)-\frac{1}{\kappa}\ln\left[\kappa(v_{H}-p(u))\right],
\end{equation}
where $\kappa>0$ is the black hole's surface gravity, and $v_H$ is the past event horizon, which is conventionally set to zero. For field modes propagating in the vicinity of $v_H$ (late time), $u\approx -\kappa^{-1}\ln\left[-\kappa p(u)\right]$, and $\omega'\gg\omega$ (extreme gravitational/Doppler red-shift), one obtains
\begin{equation}
    \left|\beta_{\omega\omega'}\right|^2\approx\frac{1}{2\pi\kappa\omega'}\frac{1}{e^{\omega/T_{H}}-1},
    \label{perfect 2d}
\end{equation}
for a perfectly reflecting point mirror, and
\begin{equation}
    \left|\beta_{\omega\omega'}\right|^2\approx\frac{\mu^2}{8\pi\kappa\omega{\omega'}^2}\frac{1}{e^{\omega/T_{H}}+1},
\end{equation}
for a semi-transparent point mirror, where $T_H=\kappa/(2\pi)$ is the analog Hawking temperature. In general, the accelerating mirror radiates along the entire worldline, but only those radiated in the late time is relevant to the analog Hawking radiation. In particular, the spectrum Eq.\eqref{perfect 2d} for a perfectly reflecting point mirror is in exact accordance with the Hawking radiation emitted by a Schwarzschild black hole. Although a sem-itransparent point mirror possesses a different spectrum due to the time-dependent and frequency-dependent reflectivity, it nevertheless has the same temperature as that of a perfectly reflecting point mirror.

As previously mentioned, practical considerations in the laboratory forces us to work in (1+3)-dimensional spacetime and a mirror with some kind of longitudinal/transverse distribution. In the case of a semi-transparent mirror, it is possible to find the corresponding analytic spectrum through perturbative approach. The result is
\begin{equation}
    \frac{dN}{d^3k}=\int d^3p \left|\beta_{\mathbf{kp}}\right|^2,
\end{equation}
where \cite{Lin2021}
\begin{equation}
\begin{aligned}
    \beta_{\mathbf{kp}}&\approx \frac{\left<\mathbf{k},\mathbf{p};{\rm out}|0;{\rm in}\right>}{\left<0;{\rm out}|0;{\rm in}\right>}
    \\
    &\approx F(\mathbf{k},\mathbf{p})\times\frac{-i\mu}{16\pi^3\sqrt{\omega_{k}\omega_{p}}}
    \\
    &\quad \times \int dt\; \gamma^{-1}(t)e^{i(\omega_k+\omega_{p})t-i(k_x-p_x)x_{_M}(t)},
\end{aligned}
\end{equation}
where $\omega_{p}/\omega_{k}$ is the incident/emitted plane wave mode frequency, respectively, and
\begin{equation}
\begin{aligned}
    F(\mathbf{k},\mathbf{p})=&\int d^2x_{\perp}H(\mathbf{x}_\perp)e^{-i(\mathbf{k}_\perp+\mathbf{p}_\perp)\cdot \mathbf{x}_\perp}
    \\
    &\times\int d\zeta f(\zeta)e^{-i(k_x-p_x)\zeta},\quad \zeta=x-x_{_M}(t),
\end{aligned}
\end{equation}
is the form factor due to the mirror's longitudinal and transverse geometry, which is independent of the mirror's motion and reflectivity.

According to particle-in-cell (PIC) simulations \cite{Liu:2020}, a mirror of square-root-Lorentzian density distribution
and a finite transverse area can generate a good quality mirror. Thus, we shall consider the case:
\begin{equation}
\begin{aligned}
    V(x)=&\frac{\gamma^{-1}(t)\left[\Theta(y+L/2)-\Theta(y-L/2)\right]}{\sqrt{(x-x_{_M}(t))^2+W^2}}
    \\
    &\times \left[\Theta(z+L/2)-\Theta(z-L/2)\right],
\end{aligned}
\end{equation}
where $W$ is the half width at half maximum of the square-root-Lorentzian distribution and $L\times L$ is the transverse area. In addition, according to the plasma density profile designed in Ref.\cite{Chen:2020}, the mirror follows the trajectory:
\begin{equation}
\begin{aligned}
t(x_{_M})
=
\begin{cases}
-\frac{x_{_M}}{v},\quad v\rightarrow 1, \;0\leq x_{_M}< \infty,
\\
-x_{_M}+\frac{3\pi}{2\omega_{p0}(1+b)}\left[\frac{1+b}{1+be^{x_{_M}/D}}-1 \right],\quad \text{else,}
\end{cases}
\end{aligned}\label{trajectory 2}
\end{equation}
where $\{\omega_{p0},b,D\}$ are positive plasma mirror parameters and time $t$ is written as a function of the trajectory $x_{_M}$. This trajectory is designed such that it approximates the black hole-relevant trajectory: $u\approx -\kappa^{-1}\ln\left[-\kappa p(u)\right]$ either (i) at the late-time $(t\rightarrow\infty)$ for any value of $b$, or (ii) in a near-uniform plasma background $(b\ll 1)$ during the entire accelerating phase. In either cases, the spectrum relevant for the analog Hawking radiation is
\begin{equation}
    \frac{dN}{d\omega_{k} d\Omega}\approx\frac{\mu^2\mathcal{F}_{L}(\mathbf{k}_{\perp})\mathcal{F}_{W}(k_x)}{8\pi\kappa}\left[\frac{\omega_k}{e^{\omega_k/T_{\rm eff}(\theta_k)}+1}\right],
\end{equation}
where $T_{\rm eff}(\theta_k)=\kappa/[(1+\cos\theta_k)\pi]$ is the effective temperature, $\kappa=1/(2D)$, $\mathcal{F}_{L}$ is a complicated form factor due to the mirror's transverse distribution which was given in Ref.\cite{Lin2021}, and
\begin{equation}
    \mathcal{F}_{W}(k_x)=4\int dp_{x}\;p_{x}^{-2}K_{0}\left(W\left|p_{x}-k_{x}\right|\right)^2,
\end{equation}
is the form factor due to the mirror's longitudinal distribution, where $K_{0}$ is the modified Bessel function of the second kind. Note that the form factor $\mathcal{F}_{L}$ contributes to the diffraction of field modes whereas $\mathcal{F}_{W}$ may enhance the production rate. 

Using the PIC simulation parameter values: $\mu=0.096$ eV, $\kappa=0.2$ eV ($D=0.5$ $\mu {\rm m}$), $\omega_{p0}=0.006$ eV, $W=0.0074$ eV$^{-1}$ (1.5 nm), $L=254$ eV$^{-1}$ (50 $\mu {\rm m}$), and $b=1$, the resulting analog Hawking temperature is $T_{\rm eff}\sim 0.031$ eV (369 K) in the far infrared regime and the number of produced analog Hawking particles per laser crossing is
\begin{align*}
    N\approx\int_{0}^{\kappa}d\omega_{k}\int d\Omega\;\frac{dN}{d\omega_{k} d\Omega}= (4.1+0.4)\times 10^{-3},
\end{align*}
where 4.1 and 0.4 correspond to the red and the blue areas in Fig.6, respectively.

Assuming a petawatt-class laser such as that in the Apollon Laser Facility in Saclay, France, that can provide 1 laser shot per minute and 8 hours of operation time per day, a 20-day experiment with a 100\% detector efficiency would give the total yield of events as
\begin{equation}
    N_{\rm detect}= (1\times 60\times 8\times 20)\times 1\times N\approx 43.
\end{equation}
It should be reminded that this value is highly idealized. Fluctuations of the physical parameters, especially that of the characteristic length of the density gradient, $D$, which we have not yet measured, would impact on the expected Hawking photon yield. 

\begin{figure}[!htb]
\centering
\minipage{0.4\textwidth}
  \includegraphics[width=\linewidth]{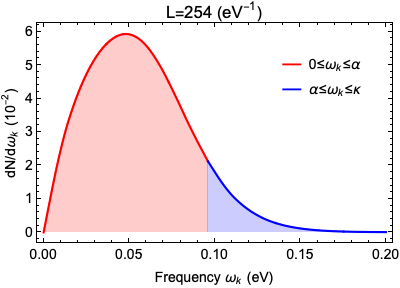}
  \caption{Frequency spectrum of analog Hawking particles. The area shaded in red gives a total number of $4.1\times 10^{-3}$ while that shaded in blue gives $0.4\times 10^{-3}$.}\label{fig:frequency spectrum}
\endminipage\hfill
\end{figure}

\section{Supersonic Gas Jet}


As estimated in Ref.\cite{Chen:2017}, the gradient of the electron number density required for the experiment is $\sim10^{20}/{\rm cm}^3/{\rm cm}$, which is attainable with a supersonic gas jet. There are several methods proposed in the literature, such as a shock wave generated induced by a laser that propagates perpendicular to the gas jet \cite{kaganovich:2014,helle:2016}, and a supersonic gas flow impinged by a thin blade \cite{shcmid:2010,FangChiangL:2020}. The estimated gradient of electron number density reached by different groups in \cite{kaganovich:2014,shcmid:2010,FangChiangL:2020} are summarized in Table \ref{table:density_gradient_summary}. It is clear that, in principle, both methods can provide gradient that satisfies our requirement. As our first attempt, we chose the latter method because for its simplicity. 

\begin{table*}[t]
\small
  \begin{center}
    \begin{tabular}{lccccc}
\hline
Method & Laser-induced shock wave &  Blade-induced shock wave & Our target value \\
\hline
Groups &  Kagonovich et al. (2014) &  Schmid et al. (2010);   Fan-Chiang et al. (2020)  &\\
\hline
 $(\frac{\partial n_e}{\partial x})_{max}$ $[{\rm cm}^{-4}]$ & $10^{22}$ &  $\sim4\times10^{22}$\quad\quad\quad;  \quad\quad\quad$\sim 10^{20} \quad\quad\quad$ & $2\times10^{20}$ \\
 \hline
    \end{tabular}
    \caption{
The maximum gradient of electron number density obtained from different groups. Our target value is also shown.}
    \label{table:density_gradient_summary
}
    \label{table:density_gradient_summary}
  \end{center}
\end{table*}

\begin{figure}[htp]
    \centering
    \includegraphics[width=8cm]{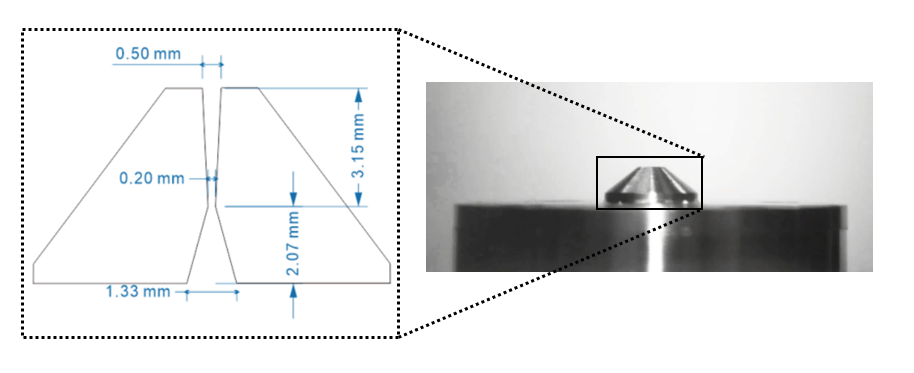}
    \caption{(left) Sketch of the nozzle used in our work. (right) The photo of our nozzle.}
    \label{fig:nozzle}
\end{figure}

\begin{figure}[htp]
    \centering
    \includegraphics[scale=0.60]{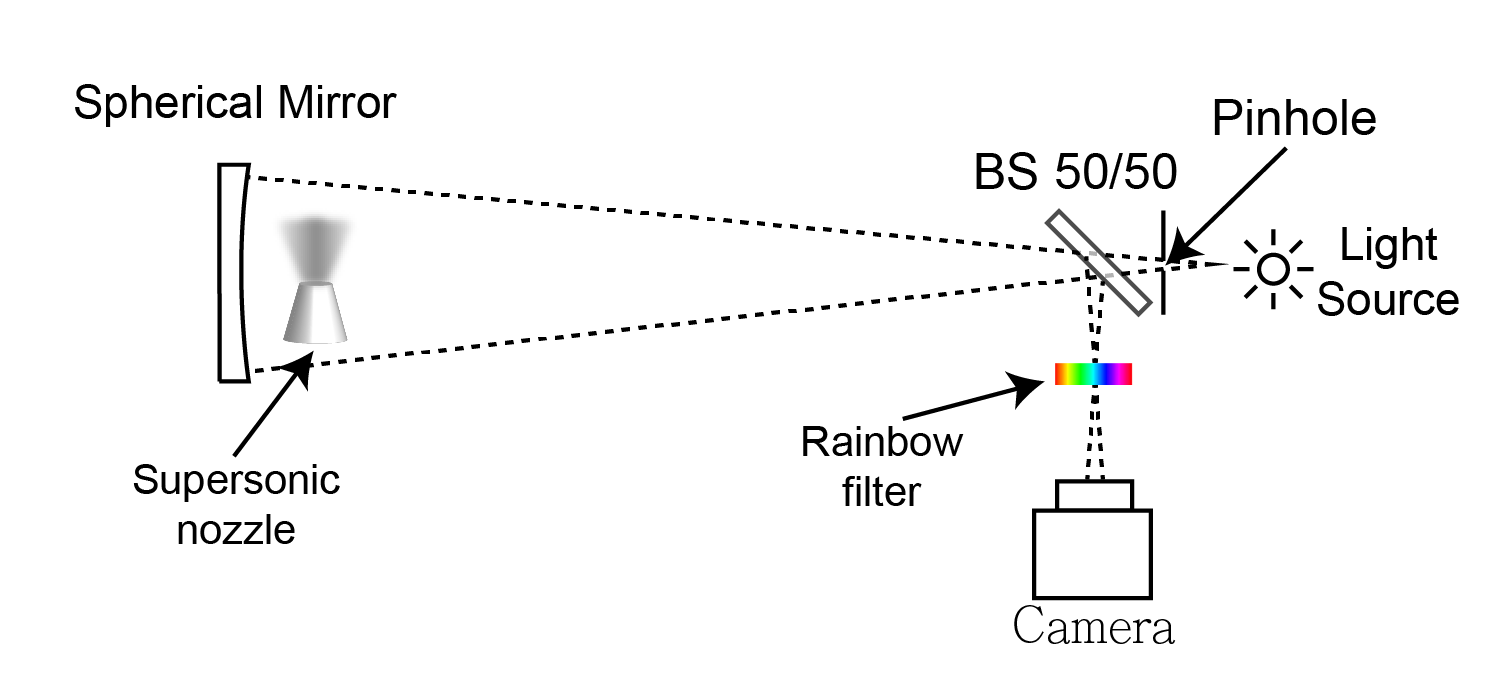}
    \caption{Schematic diagram of our Schlieren optics.}
    \label{fig:schematicSchlieren}
\end{figure}

The supersonic gas jet can be realized by passing a high pressure gas through the de Laval nozzle, which is also known as the converging-diverging nozzle. The gas flow will reach sonic speed at the throat of the nozzle and then be accelerated in the diverging section to reach supersonic speed. Based on the design of the nozzle in \cite{Hsu-hsin Chu:2005}, we produce our own nozzle to generate supersonic gas flow. Fig.\ref{fig:nozzle} shows the inner geometry and the image of the nozzle we built. The nozzle is connected to the tank of an air compressor that can provide air with pressure up to 8 atm. An electrically controlled valve is placed between the nozzle and the tank to control the flow.
There are several techniques to quantitatively characterize the density of a supersonic gas jet, including interferometry and shadowgraphy \cite{golovin:2016,kim:2018,Fang:2018,Hansen:2016}, tomography \cite{golovin:2016,Couperus:2016,Adelmann:2018}, Planar laser-induced fluorescence (PLIF)  \cite{FangChiangL:2020,Epstein:1974,Hanson:2018}, Schlieren optics \cite{settles:2001,mariani:2020} (more references can be found in \cite{mariani:2020}). As the first step, we built a Schlieren imaging system in the lab for the jet characterization. Our Schlieren optics is equipped with a rainbow filter, which allows for the visualization of the gas jet as well as quantitative analysis of its refractive index. Figure \ref{fig:schematicSchlieren} demonstrates the schematic diagram of our system.

The principle behind the Schlieren optics is that the variation of the refractive index would diffract light. A rainbow filter that intercepts the diffracted light then provides information that would quantitatively determine the diffraction angle according to the color codes. The imaging system is calibrated with a plano-convex lens whose refractive index is known. In this way, the map of refractive index gradient, which is directly related to the gas density gradient, can be obtained. 

\begin{figure}[htp]
    \centering
    \includegraphics[height=150pt]{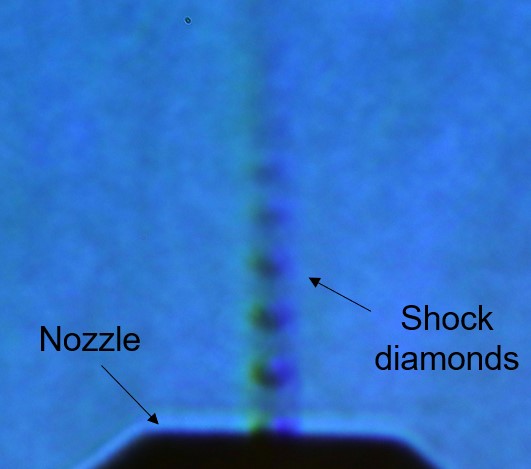}
    \caption{Obtained image with our Schlieren imaging system. Supersonic jet with shock diamonds are shown.}
    \label{fig:obtaiedImageSchlieren}
\end{figure}

Figure \ref{fig:obtaiedImageSchlieren} shows the image using our Schlieren optics. The figure shows supersonic jet produced by the nozzle. The so-called “shock diamonds” are clearly demonstrated, which is an indicator of the jet propagating with supersonic speed in the atmosphere. 

The design of the nozzle is verified by comparing the shock diamond structure from the data with the computational fluid dynamic (CFD) simulation result. The 3D fluid simulation was performed with OpenFOAM code. In the simulation, a compressible Navier-Stokes flow solver, rhoCentralFoam \cite{rhoCentralFoamSolver}, is used to study the behavior of the supersonic jet.

\begin{figure}[htp]
    \centering
    \includegraphics[height=150pt]{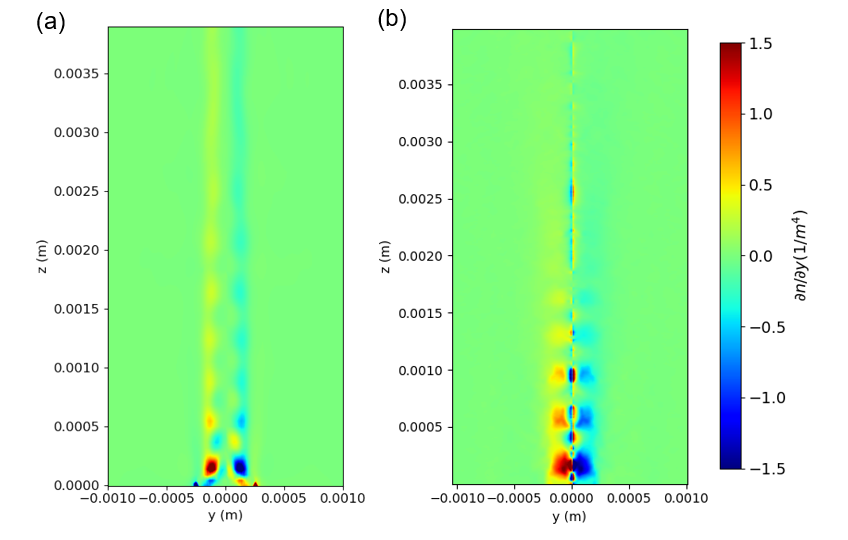}
    \caption{Two-dimensional map of the gradient of the refractive index, $\partial n/\partial y$, based on (a) Simulated result and (b) Reconstructed data. Here $y$ and $z$ are the horizontal and vertical coordinates, respectively.}
    \label{fig:refractive_index_gradient_map}
\end{figure}

\begin{figure}[htp]
    \centering
    \includegraphics[height=150pt]{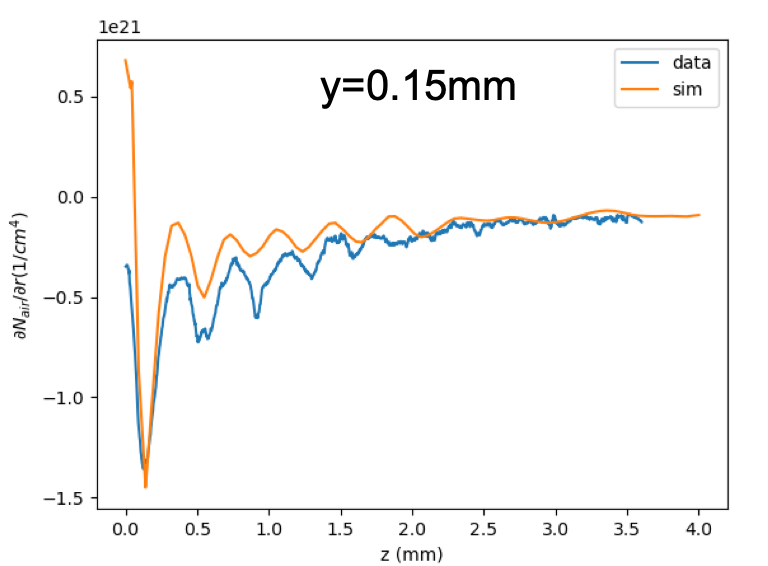}
    \caption{The line profile of $\partial n/\partial y$ as a function of vertical position $z$. The data successfully captures the position of first few shock diamonds. }
    \label{fig:density_gradient}
\end{figure}

With the conventional Abel inversion technique, the gradient of the refractive index was reconstructed and compared with the simulation result in Fig.\ref{fig:refractive_index_gradient_map}.
Line profiles at different horizontal position, $y$, relative to the axial center of the gas jet are shown in Fig.\ref{fig:density_gradient}. We found the position of several peaks in the data agree reasonably with simulation results. This implies the behavior of our self-made supersonic nozzle is as expected and our Schlieren optics can characterize the profile of the supersonic jet. Further improvement is ongoing to obtain result with higher accuracy.

\section{Superconducting Nanowire Single-Photon Hawking Detector}
Observation of {\it Hawking} photons is the main goal and one of the
major challenges of the planned AnaBHEL experiment. 
There is probably no single technology that satisfies all
requirements. The detector must be a single photon detector, with 
efficiency close to 100\%. The desired Hawking photon sensitivity 
wavelength range should be from 10 $\mu$m down to 100 $\mu$m. 
A second detector design is required for the forward moving 
{\it partner} photon with sensitivity at the UV (1-100 nm).
The low expected signal yield and the potentially large
asynchronous thermal and plasma induced backgrounds set
stringent detector timing requirements (to picosecond level 
or better). Since within the data acquisition timing window 
accidental coincidences may still be present, single photon 
pair polarization measurement will be required in order to 
unambiguously tag the pair as {\it Hawking} and {\it partner} photons.
In addition to the above requirements, 
the detector should have very fast recovery to 
avoid photon pile-up, a very low dark current rate (DCR), and the 
ability to cover relatively large areas.

Superconducting nanowire single photon detectors (SNSPDs) is the 
technology that satisfies most of the above requirements 
\cite{Zadeh}. Thin superconducting films ($\sim 10$ nm) from 
materials such as NbN and WSi are sputtered on substrates.
Subsequently, electron nanolithography is used to etch 
narrow wire structures (50-100 nm wide). The detector operates 
at a temperature below the Curie temperature $T_C$ at an 
appropriate bias current that maximizes efficiency. Additional 
cavity structures are needed in order to bring the efficiency
close to 100\%.

The intrinsic time jitter of SNSPDs is $\sim 1$ ps.
Recently, time jitters using short straight nanowires and are found to be $<3$ ps for NbN \cite{NbNrec} and 4.8ps for WSi wires \cite{WSirec}.
Thanks to their short reset time, these devices exhibit very
high count rates at the level of hundreds of MHz.
Although the expected Hawking photon yield is low, such a fast 
recovery detector reduces dramatically the probability of 
photon pileup (multiple counts in the same time window).
The dark count rate (DCR) is extremely low at the level of 
one count for a period of hours, depending on the operating 
temperature and the bias current.

Typical SNSPD designs relevant to AnaBHEL are based on a superconducting
nanowire patterned from a thin film of thickness between 5 and 10 nm.
The most common nanowire design follows a meandering structure
geometry. However, in our case we need to consider specific
structures that have sensitivity to polarization.
SNSPDs are DC-biased with an operation current close to their critical 
current so that efficiency is maximized. 
As discussed in \cite{Zadeh}, the detection
process is divided in the following steps: 
(I) photon absorption; 
(II) creation of quasiparticles and phonons combined with
their diffusion; 
(III) emergence of a non-superconducting nanowire
segment; 
(IV) re-direction of the bias current in readout circuitry,
leading to a voltage pulse; and (V) detector recovery.

During step (II), the impinging near-IR photon photo-excites an
electron and the relaxation of which leads to the formation
of a cloud of quasiparticles and phonons.
An instability of the superconducting state emerges due to the
quasiparticle cloud, which results in the reduction of the effective 
critical current density and a part of the nanowire 
experiences a transition to the non-superconducting state (III).
The occurence of a normal-conducting hot spot in the nanowire can lead
to the detection of the photon event as the current flowing through
the bias resistor (bias current) is re-directed.
Due to internal Joule heating, the resistive domain of the nanowire 
keeps growing, which leads to an increased resistance at the level of 
k$\Omega$. This significant non-zero resistivity causes the 
redirection of the bias current from the nanowire to the 
readout electronics (IV). Finally, the resistive domain is cooled 
down and the superconductivity is restored, bringing the nanowire back to its 
initial state (V).

Specific requirements of the AnaBHEL experiment photon sensors are 
summarized in Table~\ref{tab:specs} (first row). Realistic operational
parameters and performance for typical SNSPD materials are also 
presented.
\begin{table*}[t]
\small
  \begin{center}
    \begin{tabular}{lccccc}
\hline
 Material  & Curie T (K)  & Operating T (K) & wavelength ($\mu$m)  & efficiency
 [\%] & t-jitter (ps)\\
\hline
Requirements & $<10$ & 1-4             & $>10$ (for UV:1-100ns )& $>95$  & $<10$\\
\hline
NbN    & 10 & 0.8-2.1 & 1.55 & 92-98.2 & 40-106 \\
NbTiN  & 14 & 2.5-2.8 & 1.55 & 92-99.5 & 14.8-34 \\
WSi    &  3 & 0.12-2  & 1.55 & 93-98   & 150    \\
MoSi   &$<3$& 0.8-1.2 & 1.55 & 80-87   & 26-76  \\
MoSi (UV)&5 &  $<4$   & 0.250& 85      & 60   \\ 
    \hline
    \end{tabular}
    \caption{
SNSPD superconducting material properties and performance for specific
designs summarized in \cite{Zadeh}. Operating prototype WSi sensors for
wavelengths close to 10$\mu$m have been reported in \cite{Verma1}.
}
    \label{tab:specs}
  \end{center}
\end{table*}
In most applications, SNSPDs are coupled to fibers with a typical 
operation wavelength at the telecom window (1550 nm). 
AnaBHEL is an open air experiment with a tight requirement of 
operation at mid to far infrared 
($\lambda>10$ $\mu$m) regime. As reported in \cite{Verma1} and \cite{Wollman2}, 
significant progress has been made for open air longer wavelength
operating SNSPDs. To achieve sensitivity for wavelengths longer than 
10 $\mu$m, materials of lower Curie temperature must be used. WSi is 
an example of such a material. However, further R\&D on other materials 
is needed.

In addition to efficiency, successful detection of Hawking and partner photons in
AnaBHEL requires good detector acceptance in both the forward and 
backward part of the experimental apparatus. A single pixel SNSPD 
covers a very small active area of order of $10\times 10$ $\mu$m$^2$.
To maximize photon acceptance, a $1\times 1$ mm$^2$ pixel array would be
preferred. Such a kilopixel array has already been produced \cite{Wollman1}
and used in exoplanet transit spectroscopy in the mid-infrared regime.

\subsection{Hawking Photon Sensor Fabrication and Characterization}
\label{sec:fab}
In 2021, an R\&D program to develop photon sensors for Hawking photon
detection was initiated in Taiwan. Academia Sinica, NTU and NCU 
groups are currently sharing equipment and laboratories for the
fabrication and testing of prototype SNSPDs, the preferred technology 
for Hawking-photon sensors. 

\begin{figure}[htb!]
\centering
\resizebox{0.45\textwidth}{!}{
\includegraphics{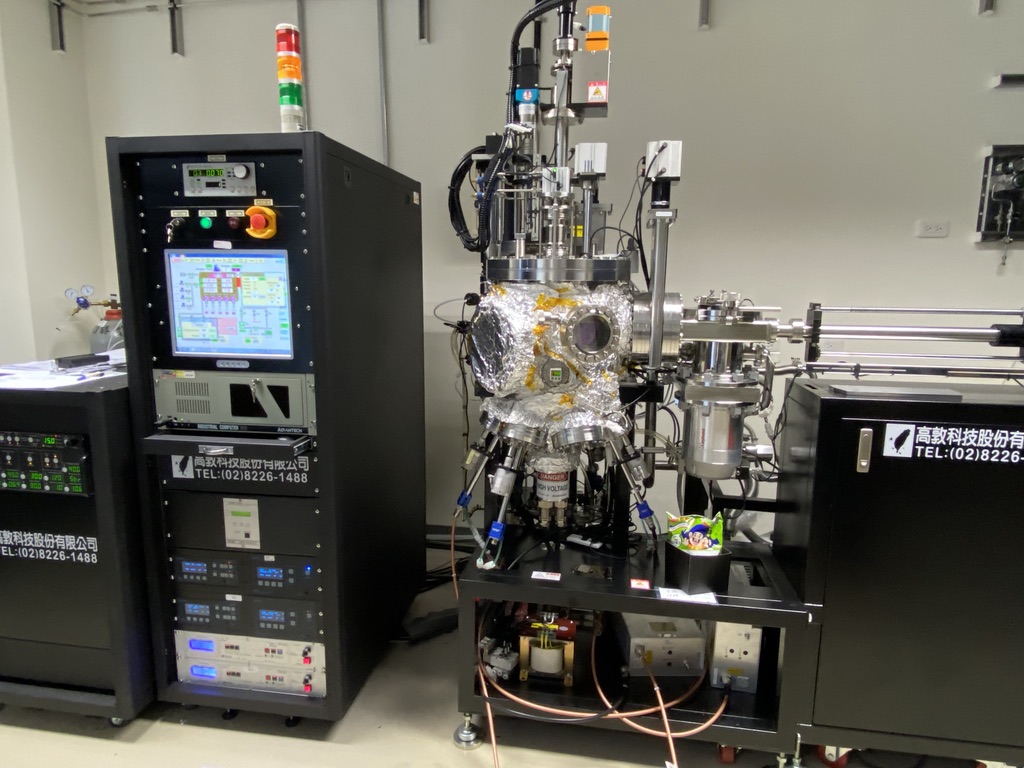}
}
\caption{Sputtering machine for film production (Kao Duen Technology, Model: KD-UHV, N-11L17)}
\label{fig:SPUT_AS}
\end{figure}

\begin{figure}[htb!]
\centering
\resizebox{0.2\textwidth}{!}{
\includegraphics{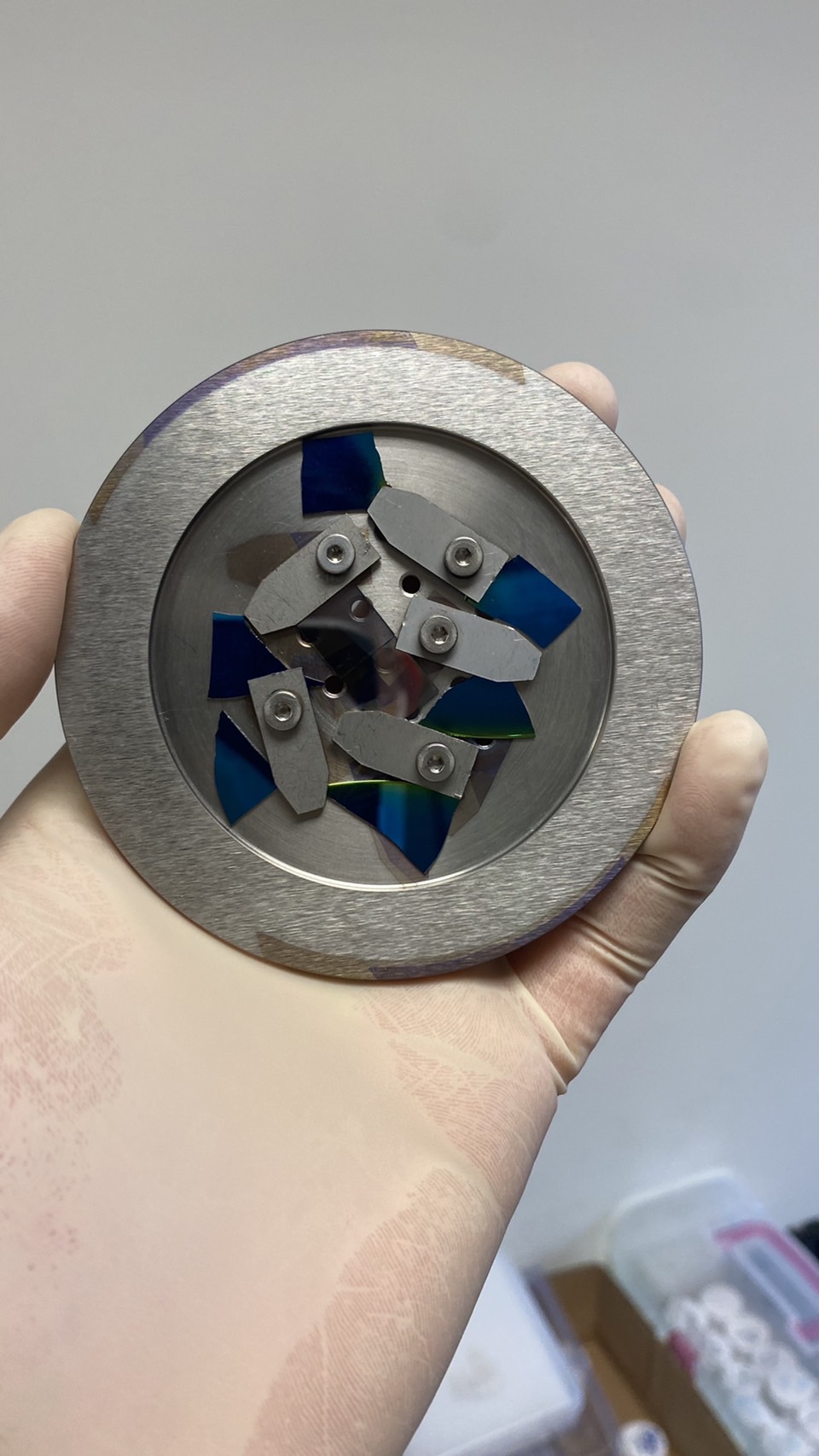}
}
\caption{NbN films on a sample holder as they come out from the
  sputtering machine. The blue sample pieces are 10nm-thick NbN on DBR substrate. The gray sample piece shown in the middle is
  10nm-thick NbN grown on MgO substrate. The difference in color is
  due to the fine NbN layer thickness.}
\label{fig:NbN_AS}
\end{figure}

We have been producing NbN films of 10~nm thickness 
using the Academia Sinica magnetron sputtering machine shown in
Fig.~\ref{fig:SPUT_AS}
(Kao Duen Technology, Model: KD-UHV, N-11L17).
The films grown on two different substrates, MgO and
distriubuted Bragg reflector (DBR), were used. The films were sputtered at UHV 
pressure of $10^{-9}$ Torr. Sample NbN films on a sample holder are shown 
in Fig.~\ref{fig:NbN_AS}.

\begin{figure}[htb!]
\centering
\resizebox{0.20\textwidth}{!}{
\includegraphics{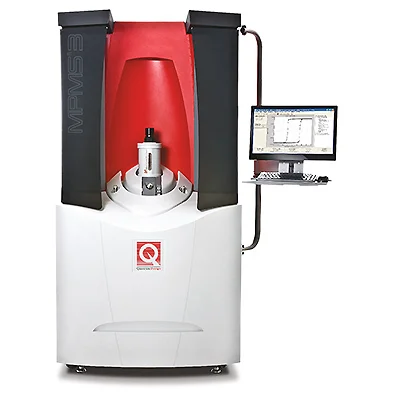}
}
\resizebox{0.27\textwidth}{!}{
\includegraphics{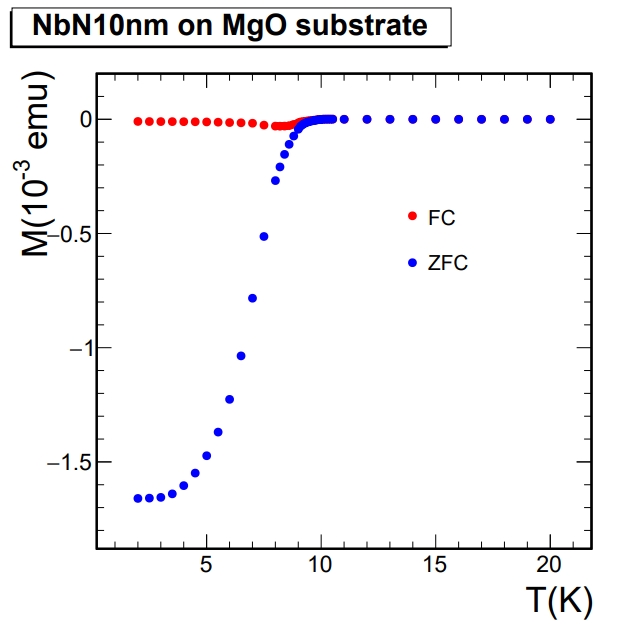}
}
\caption{MPMS3 SQUID magnetometer}
\label{fig:MS_NTU}
\end{figure}

The superconducting transition properties of the NbN films have been
determined using magnetic susceptibility measurements with a SQUID, as well as 
electric resistivity measurements. In the left side of Figure~\ref{fig:MS_NTU}, the
MPMS3 SQUID magnetometer is used to measure the magnetic susceptibility
of the NbN samples grown on MgO. In the right side of the same figure, the
superconducting transition is shown as the material
becomes diamagnetic. A $4{\rm mm}\times 4$ ${\rm mm}$ NbN sample was placed in the SQUID and
its magnetic susceptibility was measured in the temperature range of 2K-20K 
in steps as small as 0.1K per step as it approaches the Curie temperature $T_C$.

\begin{figure}[htb!]
\centering
\resizebox{0.2\textwidth}{!}{
\includegraphics{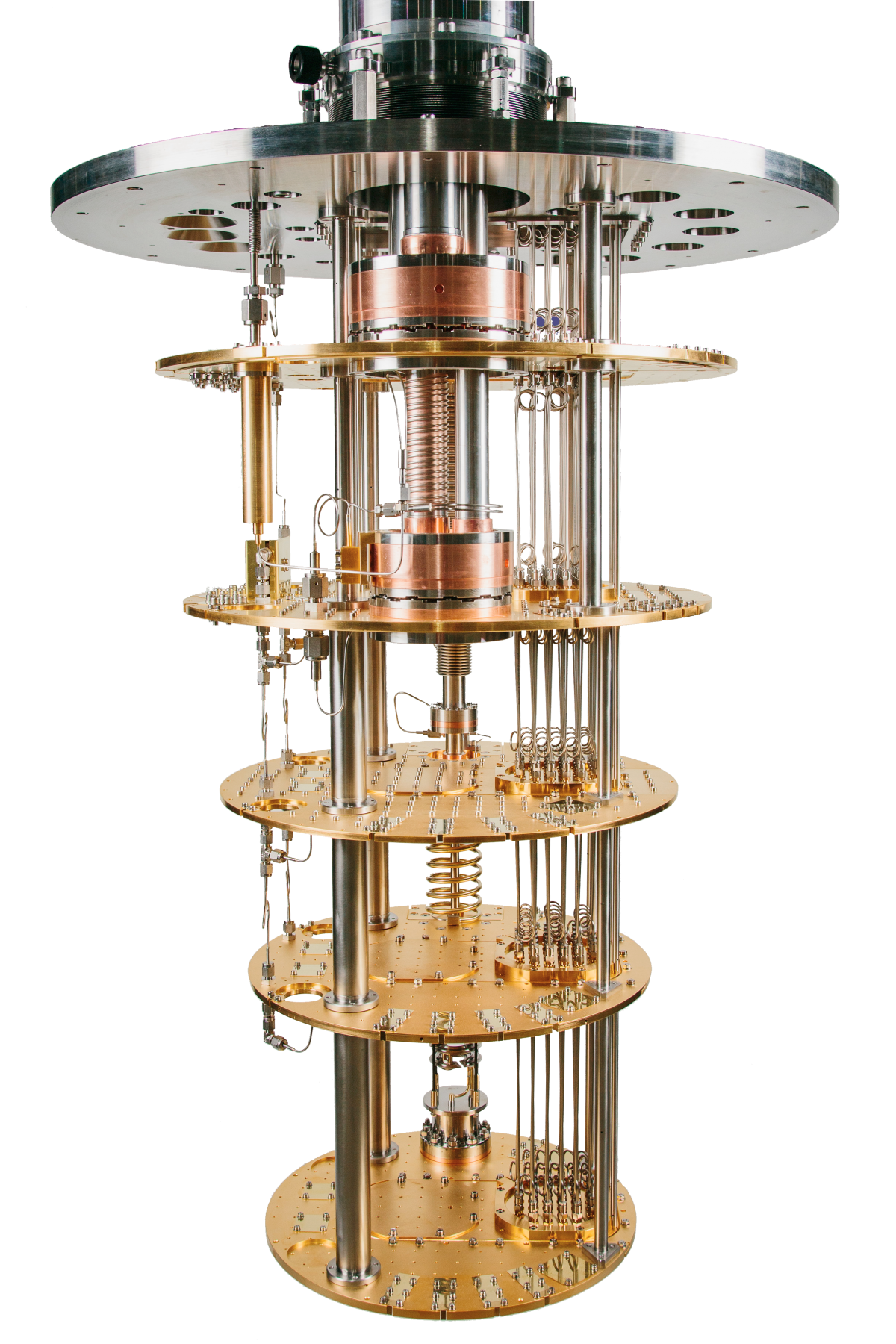}
}
\resizebox{0.27\textwidth}{!}{
\includegraphics{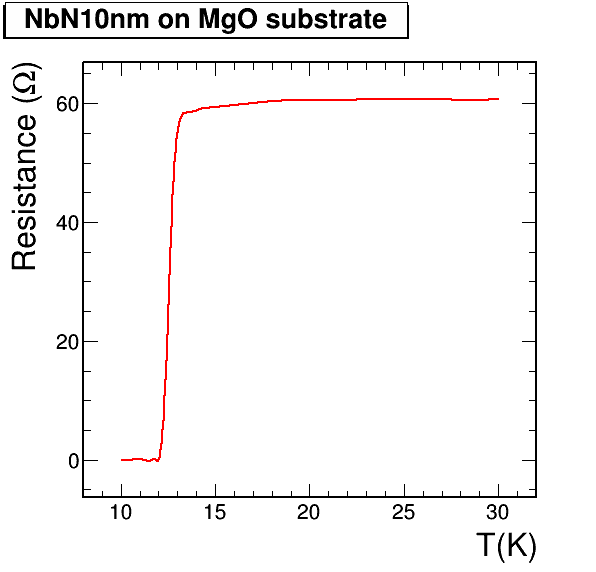}
}
\caption{Triton~500 Cryogenics setup by the NTU-CCMS group
  (left). Resistance versus temperature measurement 
of a NbN film sample grown on MgO substrate.
}
\label{fig:triton500_NTU}
\end{figure}

Electric resistivity measurements were performed with the Triton~500
cryogenics system setup by the NTU-CCMS group, shown in
Figure~\ref{fig:triton500_NTU} (left). 
A superconducting transition measurement for a NbN film sample 
is shown in Figure~\ref{fig:triton500_NTU} (right).
Samples of 3$\times$3 mm$^2$ size were prepared and glued on a sample
holder with CMR-Direct GE Varnish.
The sample was wire bonded to readout pads on the sample holder using
aluminum wires. The holder carries 20 readout pads, allowing us to
perform more than the minimum requirement of four bonds. In this way, we
ensure that we still have connectivity in case some bonds break in
very low operating temperatures.
The resistivity was first measured at room temperature to check for 
possible oxidation or defects in the film growth process, and to 
test the connectivity of the wire bonds.



\begin{figure}[htb!]
\centering
\resizebox{0.45\textwidth}{!}{
\includegraphics{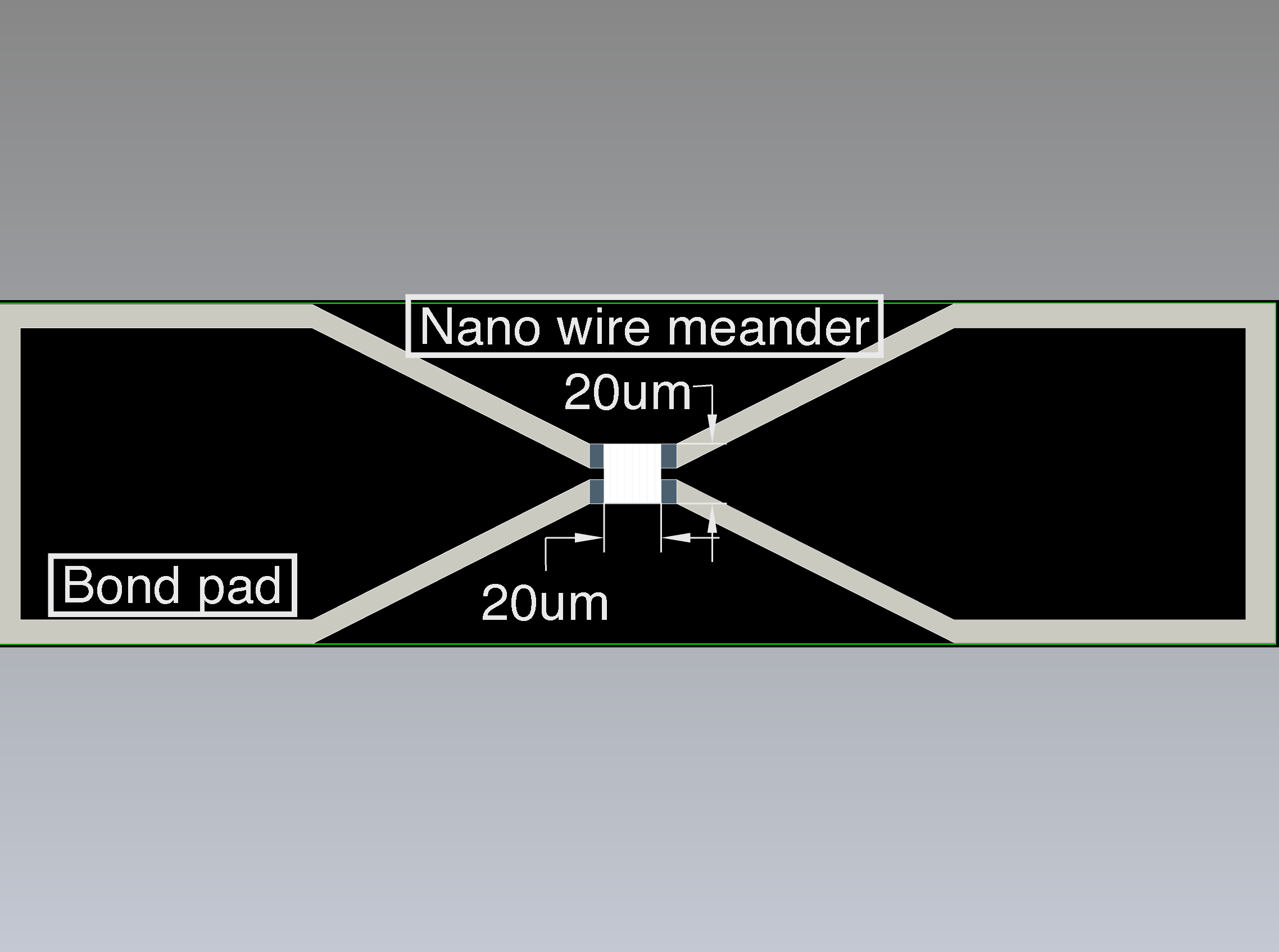}
}
\caption{Baseline SNSPD sensor prototype autoCAD drawing.}
\label{fig:SNSPD-design}
\end{figure}

After successful characterization of the NbN-sample superconducting
properties, we proceed with the production of prototype nanowire sensors.
The performance requirements for the Hawking photon sensors
necessitate the use of electron-beam lithography (EBL) for the etching 
of nanowires from the NbN films. Currently, nanowire prototypes of different 
widths and lengths are under design. The baseline design using an autoCAD drawing of a 
20$\times$20 $\mu$m$^2$ sensing-area, with nanowire with the width of 100nm and 
a pitch of 100 nm, is shown in Figure~\ref{fig:SNSPD-design}.

\begin{figure}[htb!]
\centering
\resizebox{0.45\textwidth}{!}{
\includegraphics{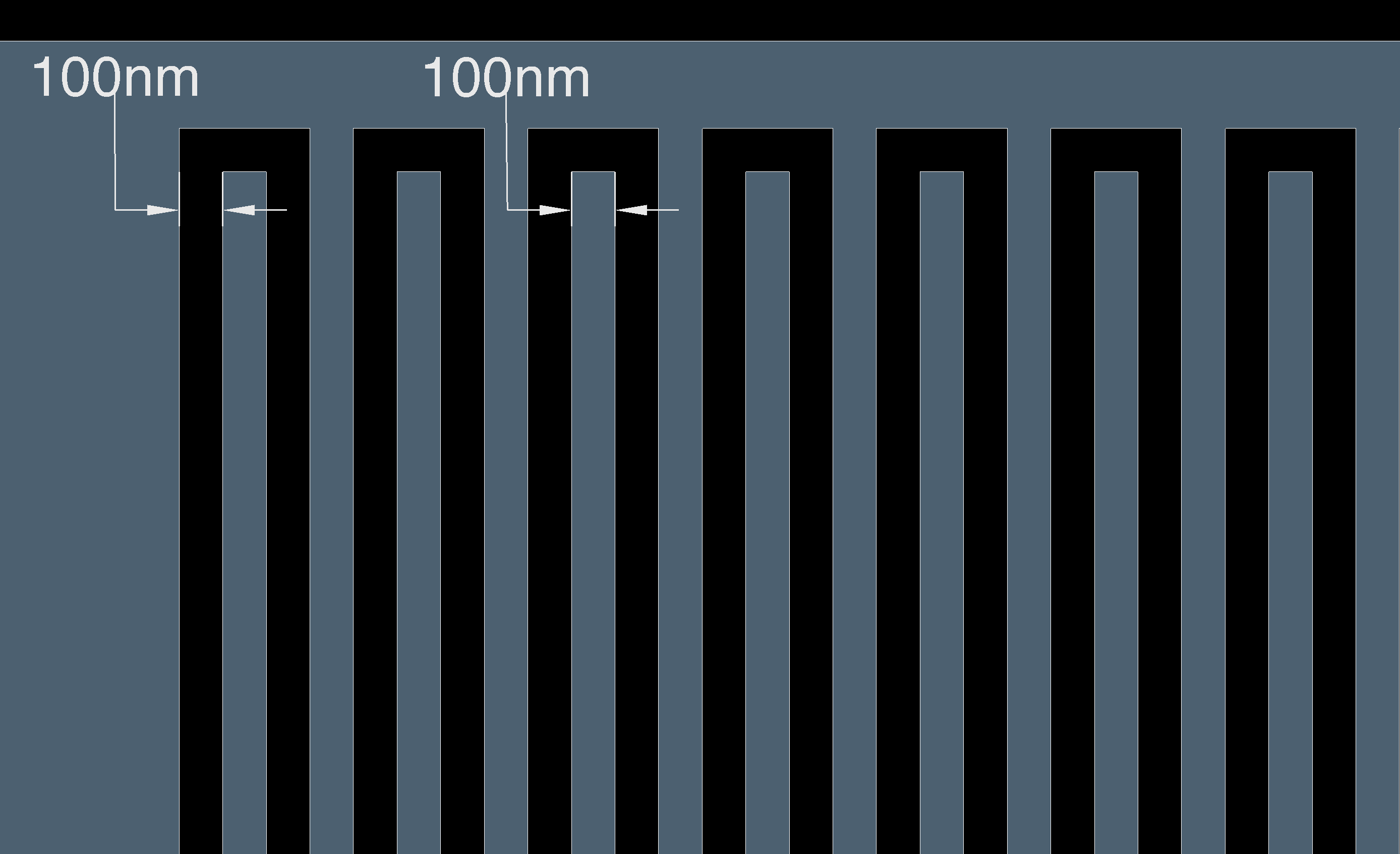}
}
\caption{Nanowire prototype autoCAD zoom-in.}
\end{figure}

The SNSPD sensor prototypes are produced by the 
EBL, ELS-7000 ELIONIX, machine located in the Academia Sinica laboratories.
\begin{figure}[htb!]
\centering
\resizebox{0.45\textwidth}{!}{
\includegraphics{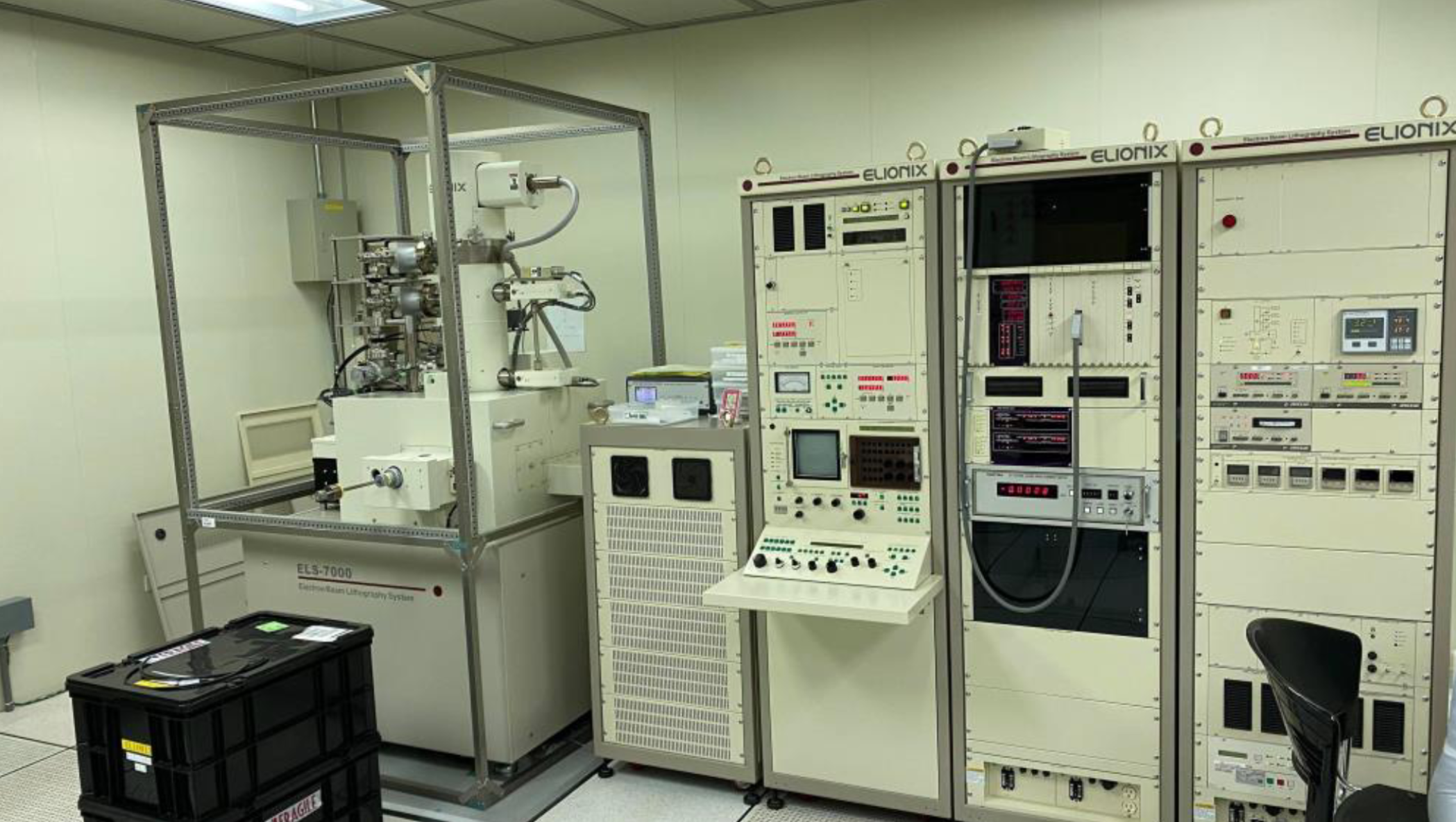}
}
\resizebox{0.30\textwidth}{!}{
\includegraphics{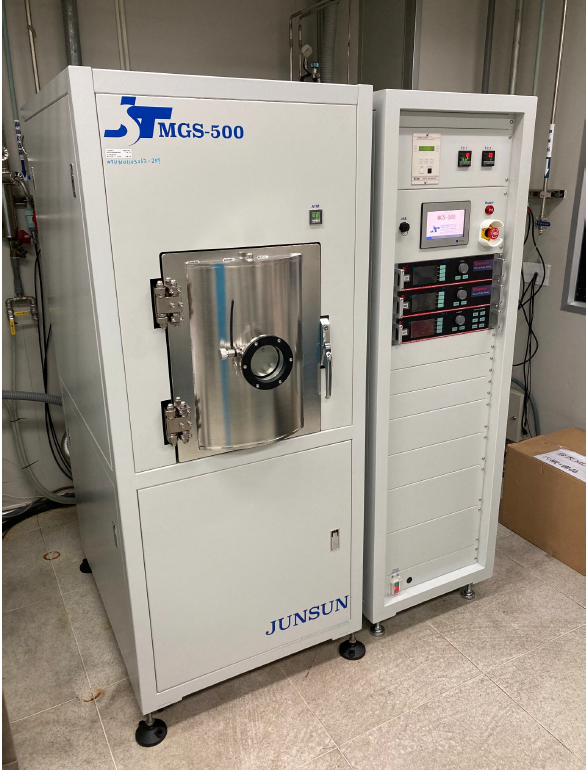}
}
\caption{Electron beam lithography machine located in (a) Academic Sinica
  laboratories (EBL, ELS-7000 ELIONIX); and (b) Junsun Tech MGS-500 sputtering machine installed at 
the NEMS center of NTU.}
\label{fig:EBL_AS}
\end{figure}
In order to maximize the single photon detection efficiency,
various structures such as cavities or Bragg reflectors can be
utilized. As part of an ongoing R\&D, distributed Bragg reflectors
(DBR) have been grown in the NTU MEMS facility.
The measured reflectivity of a DBR for a sensor sensitive at 1550 nm
is shown in Figure~\ref{fig:reflectivity}. The good agreement with a
finite-difference time-domain method (FDTD) simulation of the
structure, gives us confidence to proceed with new cavity
designs optimal for longer wavelengths.
\begin{figure}[htb!]
\centering
\resizebox{0.45\textwidth}{!}{
\includegraphics{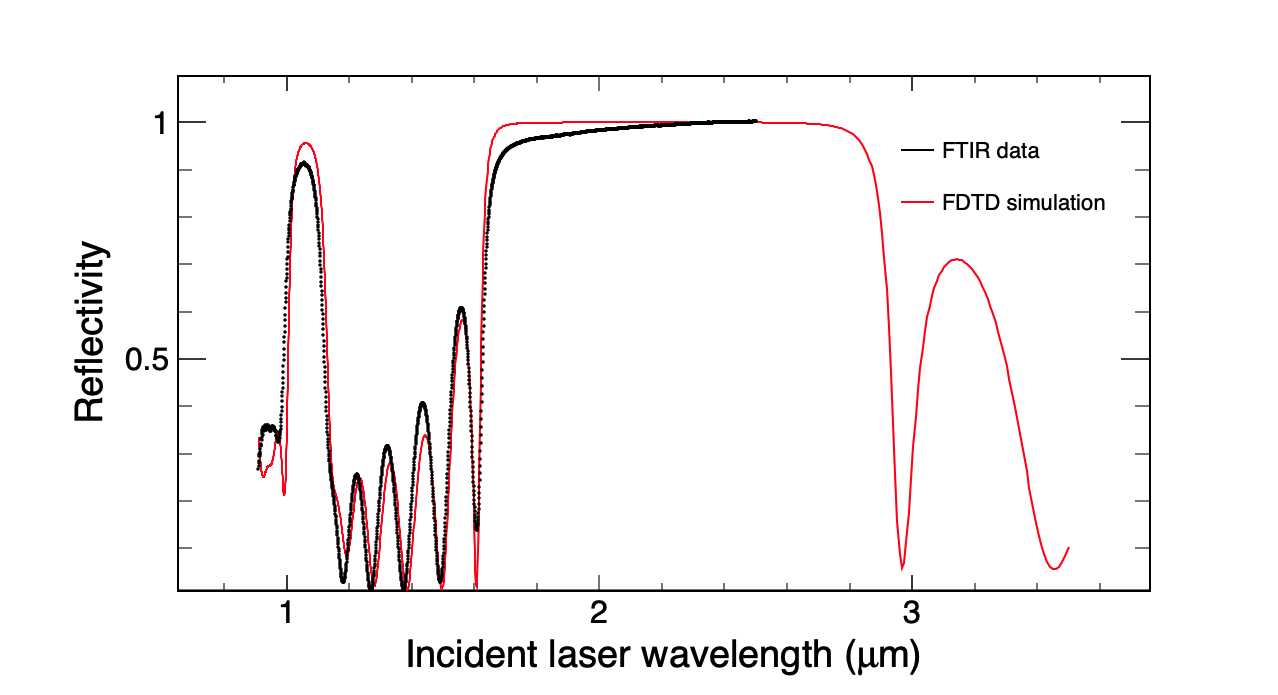}
}
\caption{Reflectivity of a distributed Bragg reflector structure used
  to enhance the efficiency of sensors sensitive at 1550 nm. Data are
  shown in black and FDTD simulation in red.}
\label{fig:reflectivity}
\end{figure}

\begin{figure}[htb!]
\begin{center}
\includegraphics[width=.9\linewidth]{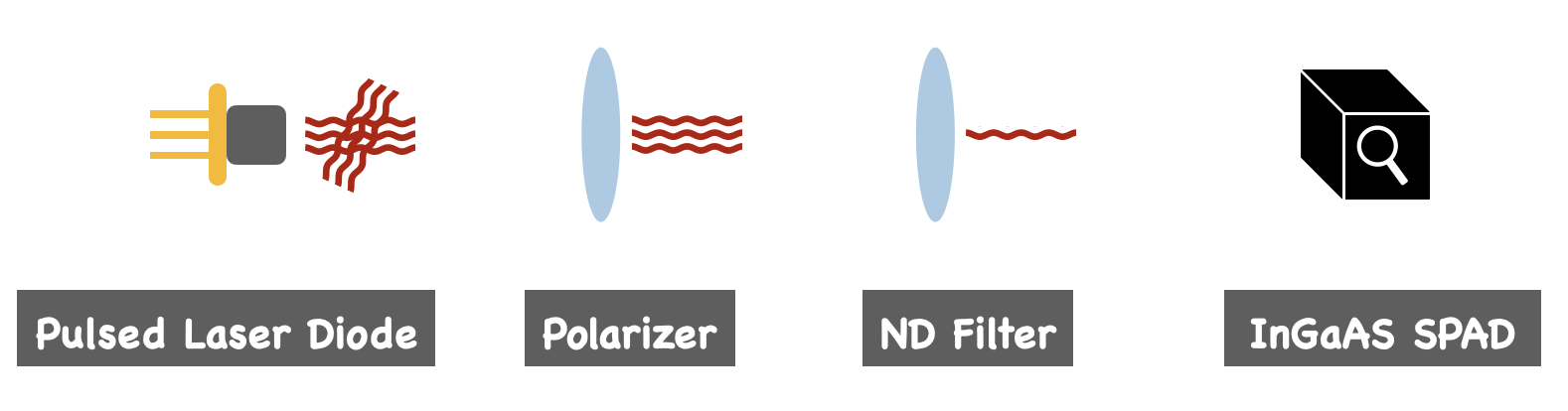}
\caption{Single photon calibration setup using SPAD.}
\label{fig:SPAD_LeCosPA}
\end{center}
\end{figure}

\begin{figure}[htb!]
\begin{center}
\includegraphics[width=.9\linewidth]{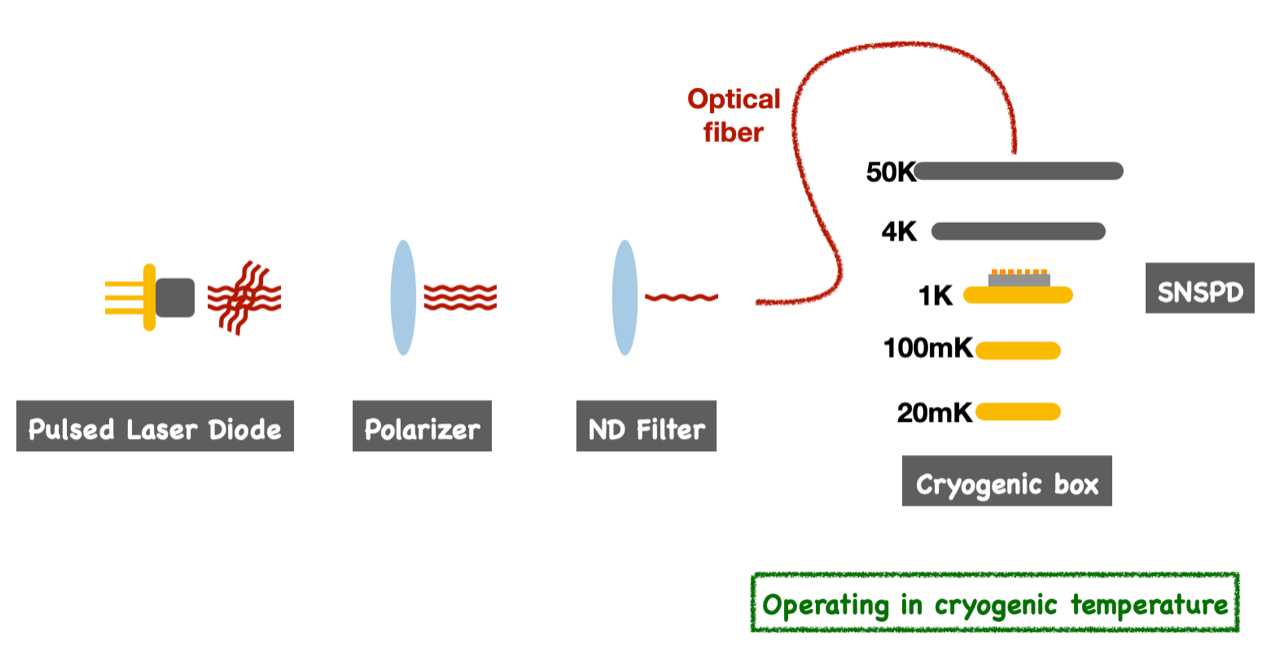}
\caption{Test-bench setup for testing sensors at 1550 nm using fibers
  connected to sensors.}
\label{fig:1550nm_LeCosPA}
\end{center}
\end{figure}

\begin{figure}[htb!]
\begin{center}
\includegraphics[width=.9\linewidth]{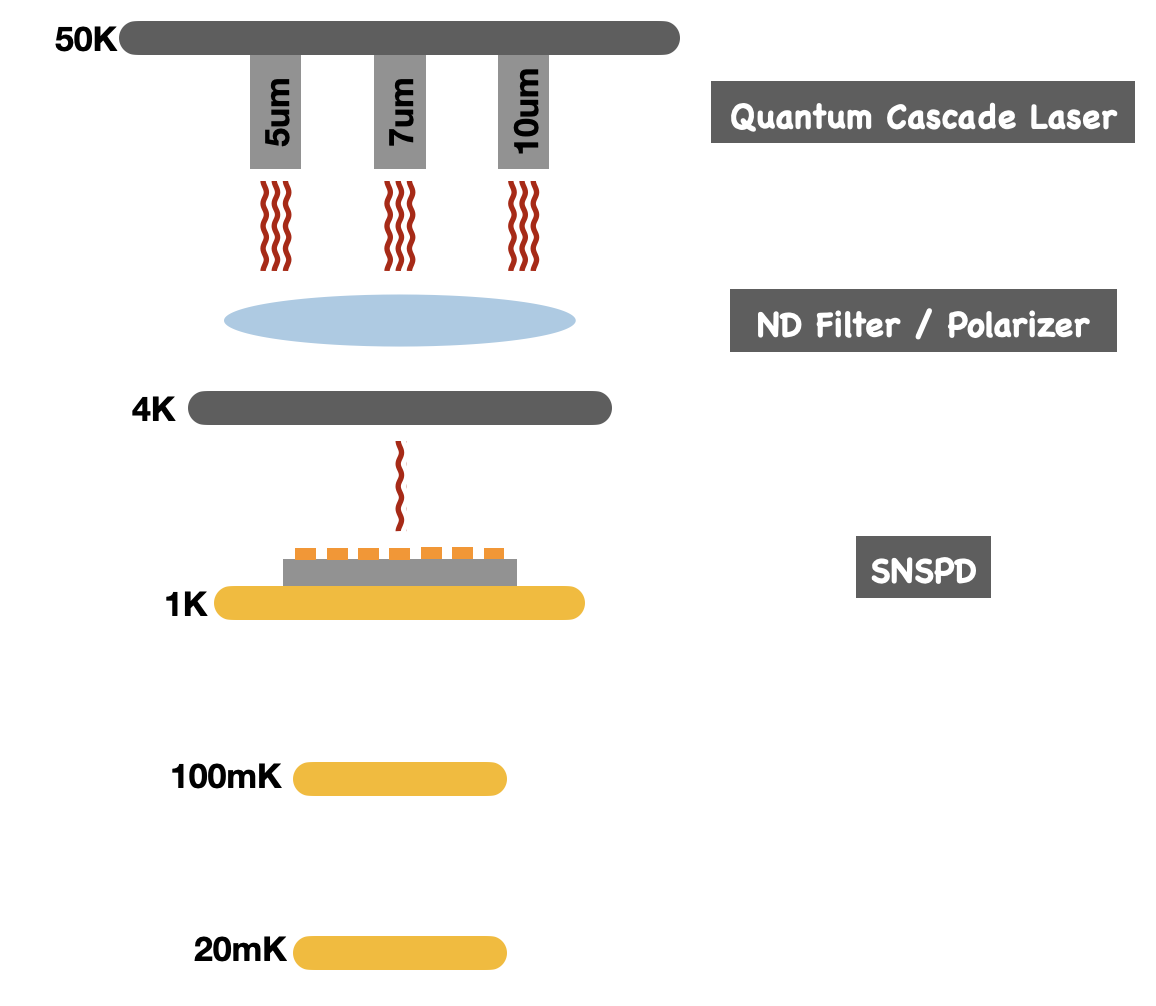}
\caption{Test-bench setup for testing sensors in open air transmission
by bringing the lasers in the cryostat. In the actual AnaBHEL experiment, the entire experimental chamber would be embedded in a cryogenic system with high vacuum.}
\label{fig:QCL_LeCosPA}
\end{center}
\end{figure}

We are currently in the process of setting up a system test bench to
characterize the Hawking sensors, using single photon at the infrared.
The setup includes a SPAD commercial sensor for single photon
calibration shown in Figure~\ref{fig:SPAD_LeCosPA}.
We plan to first verify the sensor operation at the 1550 nm where 
most commercially available SNSPDs operate, shown in Figure~\ref{fig:1550nm_LeCosPA}.
Finally, the sensors will be tested at longer wavelengths relevant 
to the AnaBHEL experiment shown in Figure~\ref{fig:QCL_LeCosPA}.
{\Large }

\section{Experimental Backgrounds}

The propagation of the high-intensity laser through a plasma target would necessarily induce background photons that would compete against the rare Hawking signals. 
The plasma electrons perturbed by the propagating laser would execute non-trivial motions and can therefore emit photons.
In addition, they can interact with the electromagnetic fields induced by the laser and charged particles, and also with the plasma ions through scatterings.

The radiations induced from interactions between the electrons and the background ions can be categorized into Thomson/Compton scattering and Bremsstrahlung.
These processes have long been well studied and the radiation so induced can be estimated when the electron trajectories are given.

There is also the possibility of radiation caused by the electron acceleration.
The analytic solution for plasma accelerating in the blowout regime of plasma wakefield excitations has been studied by Stupakov \cite{Stupakov}, where it was shown that there are not only accelerated plasmas inside the bubble, but also charged particles that oscillate along the boundary of plasma bubble.
The work \cite{Stupakov} was for the case of Plasma Wakefield Accelerator (PWFA) \cite{Chen:1985}, but the method can also be applied to Laser Wakefield Accelerator (LWFA) \cite{Tajima:1979}, which is the basis of our flying plasma mirror.
Thus we also expect to have the same type of electron motions that are oscillating around the plasma bubble.
These electrons in plasma wakefields perform a figure-8 motion in the plasma, and they can emit low-energy photons through synchrotron radiation.
These photons are propagating in the direction parallel to the laser, which could affect the observation of the partner photons downstream. 
Therefore, we should study these electrons carefully.

In the following, we categorize the trajectories of the plasma electrons obtained from simulations by using a machine-learning based technique.
We classify the electrons into several categories, according to their characteristic motions.
After this classification, we are able to identify the leading radiation processes for the electrons, and to evaluate the radiation spectrum.
We use SMILEI \cite{Smilei} for particle-in-cell (PIC) simulations and python and the scikit-learn library \cite{scikit-learn} for clustering analysis.

\subsection{Simulation Setup}
The PIC simulations are in 2D and we refer to the coordinate as $x$ and $y$.
The simulation box size is 250 $\mu \text{m} \times$ 150 $\mu \text{m}$, i.e. $0 \leq x \leq 250$ $\mu \text{m}$, -75 $\mu \text{m} \leq y \leq$ 75 $\mu \text{m}$, with $4000 \times 400$ grids.
A Gaussian laser with $800$nm wavelength and $a_0 =5.0$ is applied at the boundary of $x=0$ (left end of the simulation) and traveling in the $x$-positive direction.
We place helium gas in the simulation box that can be ionized by the impinging laser. 
The helium density $\rho_{\text{He}}$ is given by
\begin{align}
    \rho_{\text{He}} &= \begin{cases}
        \frac{n_0}{2} ( 1 + e^{- (x - \ell_0)/2)})^2, & x \geq \ell_0, \\
        2 n_0, & \text{else},
    \end{cases}
\end{align}
where $n_0 = 1$ $\text{mol} / \text{m}^3$ and $\ell_0 = 10$ $\mu \text{m}$.
We do the simulation for 265 time steps, where each step is $3.82$ femtosecond in real time.

\subsection{Categorization of Electron Motions}
Following the categorization technique introduced in \cite{Classification}, we identify electron trajectories that would induce photons that dominate the background signals. The trajectory categorization introduced in \cite{Classification} is essentially clustering in momentum space using k-mean clustering method.

Let us denote the $i$-th particle's trajectory by $p_i(t) = (x_i(t), y_i(t))$, where $x_i(t)$ and $y_i(t)$ are the $x$, $y$ coordinate of the $i$-th particle at time $t$, respectively.
The total time steps of the simulation is denoted as T. (In this case, T=265).
If we have $N$ particles to track, then our data set $\mathcal{S}$ will be $\mathcal{S} = \left\{ p_i \middle| i \in 1 \ldots N \right\}$.
Let us denote the Fourier coefficient of $x_i(t)$ and $y_i(t)$ as $\tilde x_i(k)$ and $\tilde y_i(k)$, respectively.
The categorization will be done with the following steps.
\begin{enumerate}
    \item Restrict the tracked particles data to those who has been simulated more than 380 femtosecond.
    \item \label{define dataset} Prepare a data set,
 \begin{equation}
\begin{aligned}
        \tilde S = \{ (\tilde x_i(k_1), \tilde x_i(k_2), \ldots, \tilde x_i(k_T), \tilde y_i(k_1), \ldots, \tilde y_i(k_T), \\
        \bar y_i, \bar p_{i x}, \bar p_{i y}, a_y^\text{max}, a_y^\text{min}), \quad i \in 1, \ldots, N \},
    \end{aligned}
    \end{equation}
where $\bar p_{i x}$ and $\bar p_{i y}$ are the mean of momentum of the $i$-th particle in $x$ and $y$ direction, respectively, $\bar y_i$ is the mean of $y$-coordinate of the $i$-th particle, $a_y^\text{max}$ and $a_y^\text{min}$ are the maximum and minimum of the acceleration in $y$-direction, and $k_t$ is the corresponding frequency of the $t$-th Fourier coefficient.
    \item Calculate $k$ principal component values (PCVs) from the data set. This reduces the space of clustering from $2T + 5$ dimensional vector space to $k$ dimensional vector space.
    \item Perform k-mean clustering in the $k$-dimensional space, for a given number of clusters $K$.
\end{enumerate}
Our choice of data set at step \ref{define dataset} is different from the one used in \cite{Classification}, where we have additional value $\bar y_i$ and the information of its acceleration in $y$-direction.
We added these since the longitudinal behavior is quite important for the experimental purpose, and indeed, by adding these we were able to separate the modes into more reasonable categories.

\subsection{Classification Results}
We have used $k=30$, $K=12$ in the following, i.e., we classify the particles into 12 sets by using 30 PCs. 
Although we have classified them into 12 categories, since we have included the mean value of $y$ coordinate, $\bar y_i$, into the data, we get pairs of categories that are almost symmetric along $y=0$. 
In the following, we classify those two into the same category, since their physical process are the same.

\subsubsection{1. Wakefield Accelerated Electrons}
\begin{figure}[htb]
    \centering
    \includegraphics[width=\linewidth]{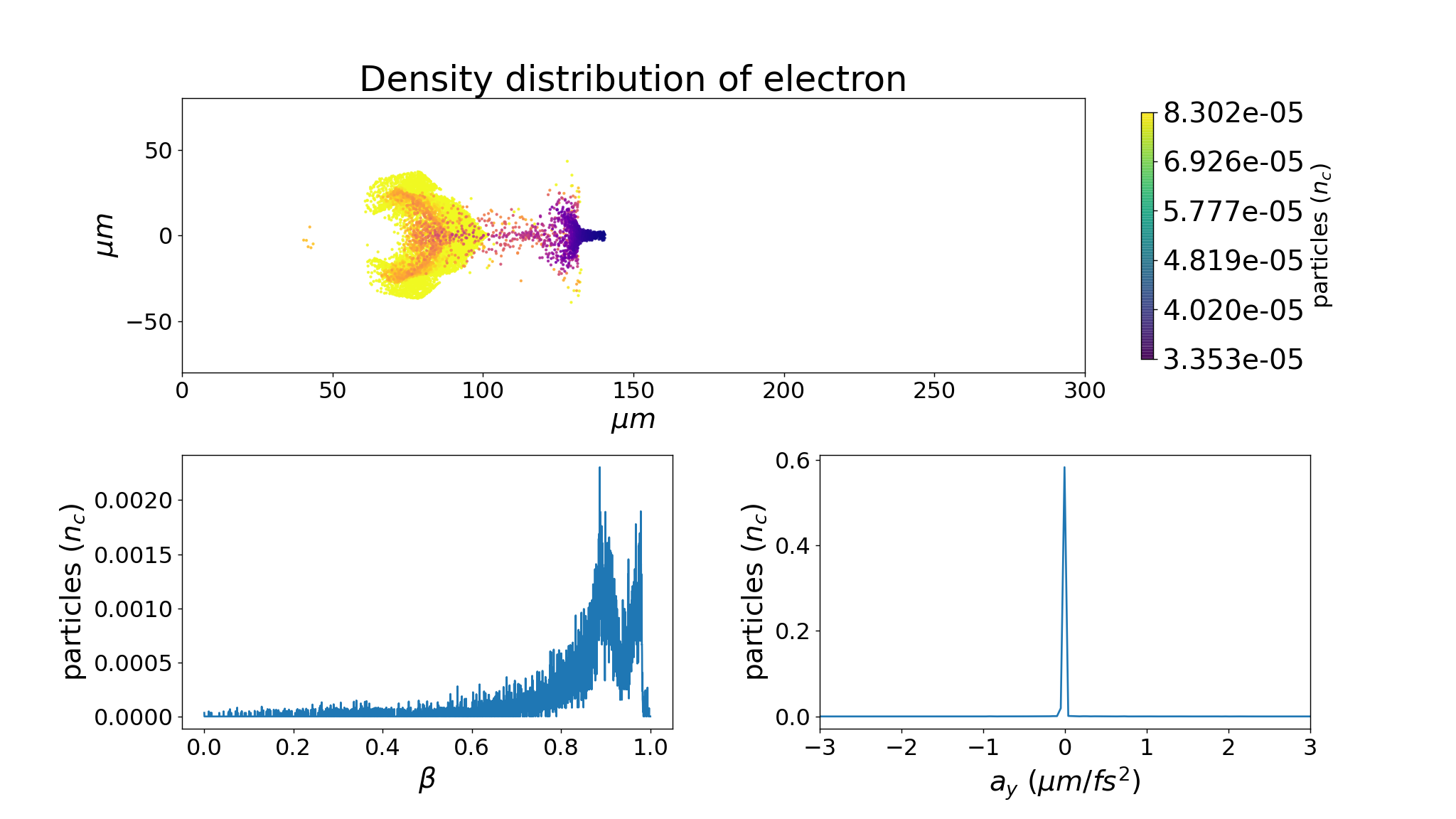}
    \caption{Snowplow electrons and electrons accelerated with the Laser Wake Field Acceleration mechanism. The top figure is the electron density distribution shown by color. Bottom left figure is the velocity distribution in $\beta = v/c$, where $v$ is the velocity of electron and $c$ is the speed of light. Bottom right figure is showing the acceleration of the electron in $y$-axis.}
    \label{LWFA electrons}
\end{figure}

The first kind is the electrons accelerated with the LWFA process. 
They are accelerated in the forward direction, to a highly relativistic regime $\beta \sim 1$.
These are shown in Figure \ref{LWFA electrons}.
These electrons can radiate photons through interacting with the nuclei, i.e., through Thomson/Compton scattering or as Bremsstrahlung.

\subsubsection{2. Snowplowed Electrons}

Snowplowed electrons are the ones that are pushed forward by the laser's ponderomotive potential and are clustered at the front of the laser pulse. 

\begin{figure}[htb]
    \centering
    \includegraphics[width=\linewidth]{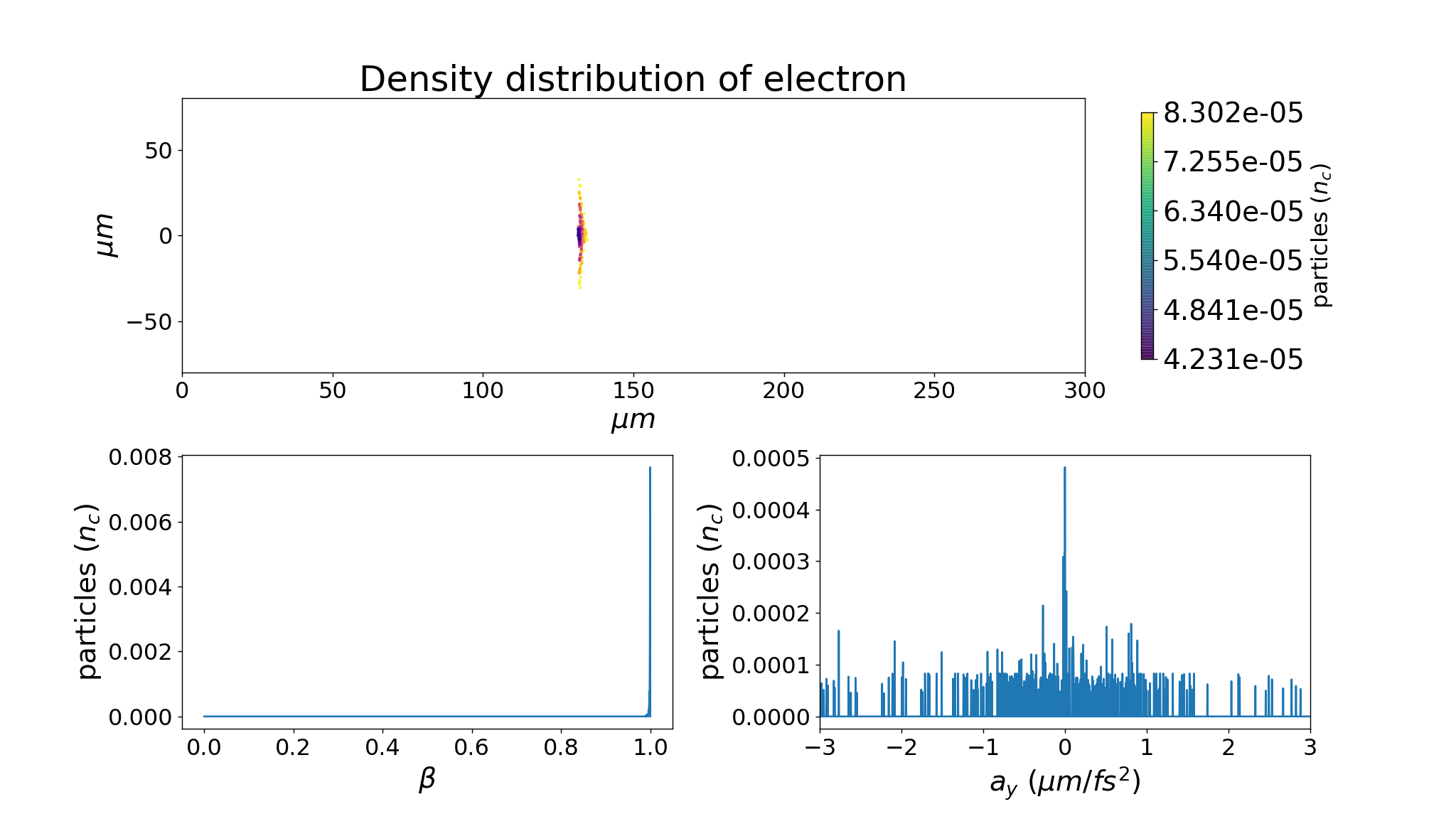}
    \caption{Snowplowed electrons. The top figure is the electron density distribution shown by color. Bottom left figure is the velocity distribution in $\beta = v/c$, where $v$ is the velocity of electron and $c$ is the speed of light. Bottom right figure is showing the acceleration of the electron in $y$-axis.}
    \label{snowplow}
\end{figure}

Figure \ref{snowplow} is a snapshot of the snowplowed electrons.

\subsubsection{3. Backward Scattered Electrons}

\begin{figure}[htb]
    \centering
    \includegraphics[width=\linewidth]{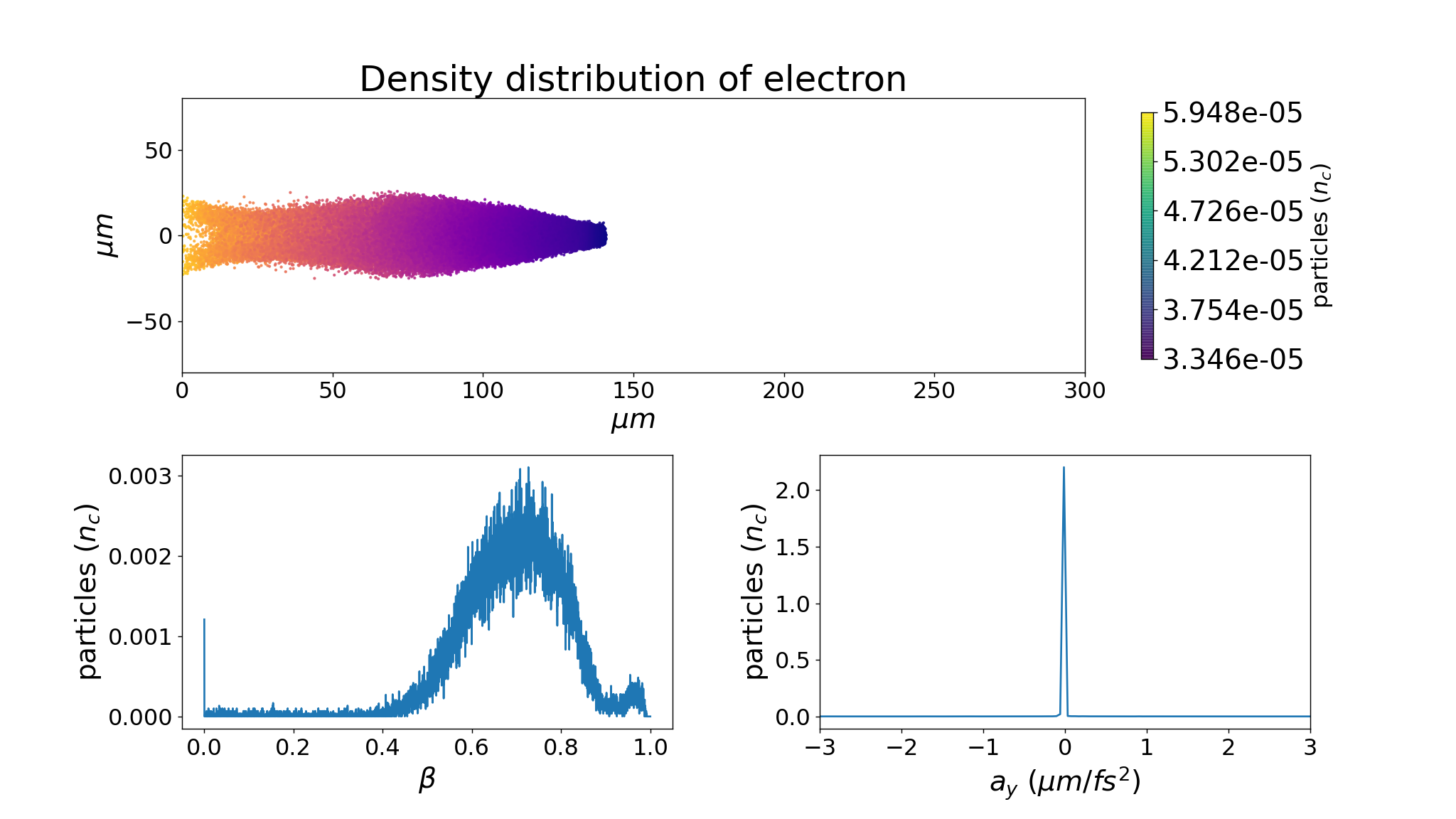}
    \caption{Backward accelerated electrons. The top figure is the electron density distribution shown by color. Bottom left figure is the velocity distribution in $\beta = v/c$, where $v$ is the velocity of electron and $c$ is the speed of light. Bottom right figure is showing the acceleration of the electron in $y$-axis.}
    \label{backward}
\end{figure}

These electrons have typically $\beta  \sim 0.7$, and are shown in Figure \ref{backward}.
They might contribute to the background radiation via Thomson/Compton scattering or Bremsstrahlung.

\subsubsection{4. Slide-away Electrons}
There are a certain fraction of plasma electrons that are pushed by the transverse plasma wakefields and propagate in the transverse direction.
In practice, they would not affect the experiment since they are not moving towards the sensor, however, one would have to consider their hitting and reflection from the gas nozzle, which would induce background photon events.
\begin{figure}[htb]
    \centering
    \includegraphics[width=\linewidth]{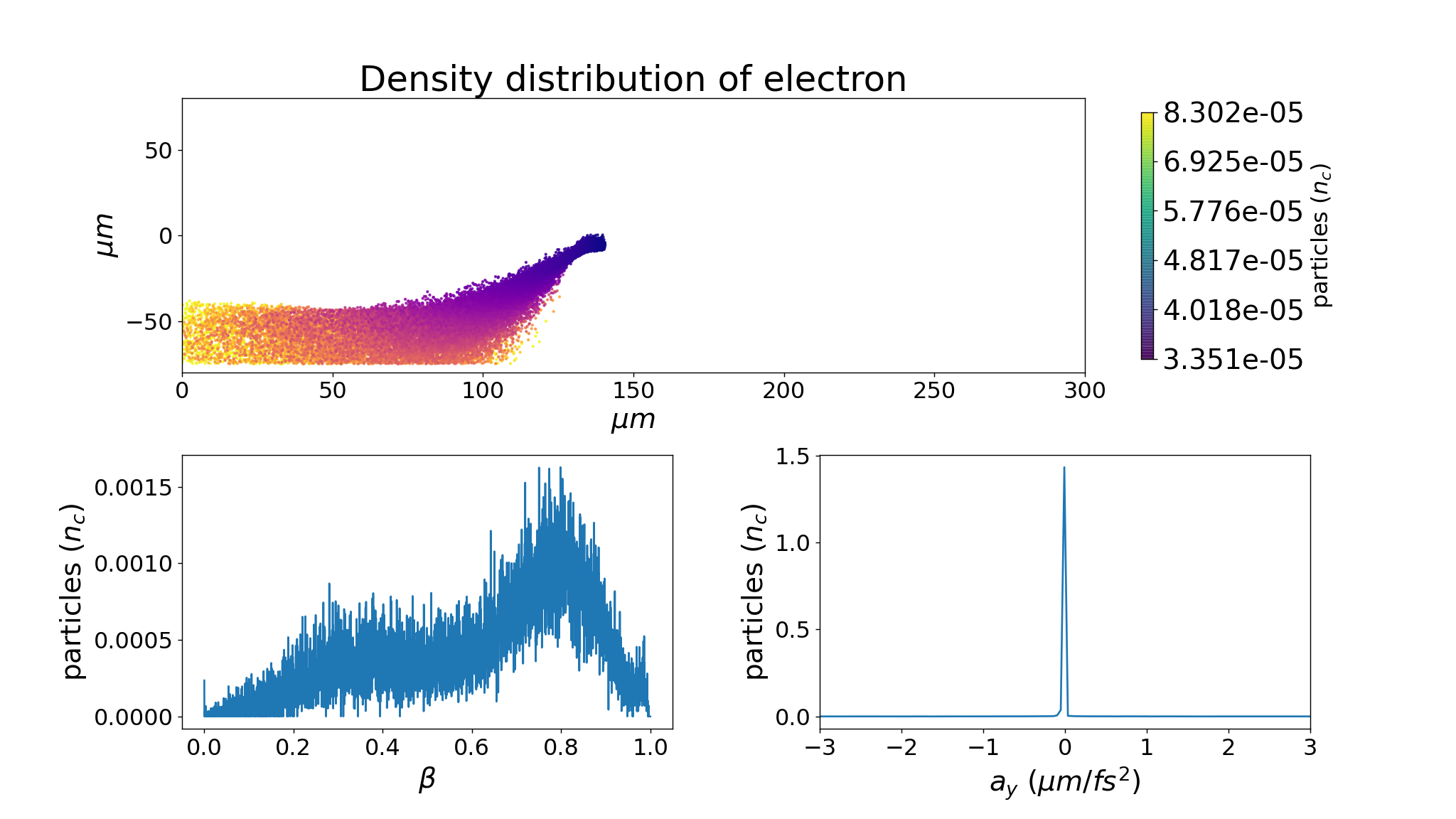}
    \caption{Slide-away electrons. The top figure is the electron density distribution shown by color. Bottom left figure is the velocity distribution in $\beta = v/c$, where $v$ is the velocity of electron and $c$ is the speed of light. Bottom right figure shows the acceleration of the electron in the $y$-axis. }
    \label{slideaway}
\end{figure}
Figure \ref{slideaway} is a snapshot of slide-away electrons. 
As pointed out previously, there are also slide away electrons that are sliding towards the positive $y$ direction, which are classified into a different category through the process.

\subsubsection{5. Transverse Oscillating Electrons}
\begin{figure}[htb]
    \centering
    \includegraphics[width=\linewidth]{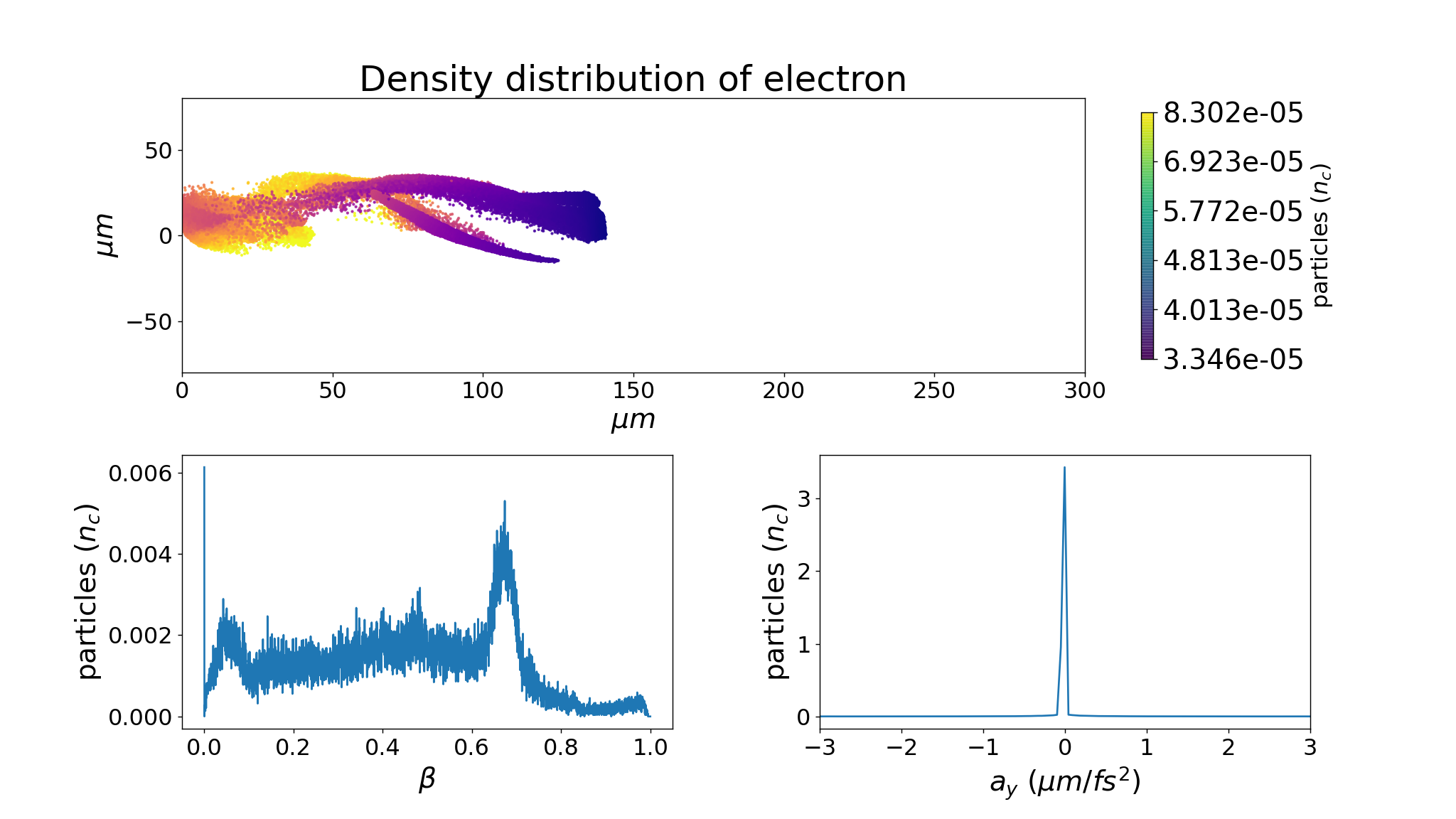}
    \caption{Oscillating electrons. The top figure is the electron density distribution shown by color. Bottom left figure is the velocity distribution in $\beta = v/c$, where $v$ is the velocity of electron and $c$ is the speed of light. Bottom right figure shows the acceleration of the electrons in $y$-axis.}
    \label{oscillating}
\end{figure}

The last one is the oscillating electrons.
These are the electrons that are attracted by the Coulomb force of the plasma ion bubble and oscillating around the laser trajectory in the traverse direction.
Figure \ref{oscillating} shows the density distribution, velocity, and acceleration in the $y$-direction of these electrons.
This distribution has a tail that extends into the non-relativistic region.
We expect that they would emit photons through synchrotron radiation, which can affect the identification of the Hawking photon signals.

\subsubsection{6. Low-Frequency Soliton Radiations}
We note that laser-plasma interaction also induces additional low-frequency background photons emitted by collective effects such as solitons, which are not included in the above discussion. Such near-plasma-frequency radiation was first pointed out by Bulanov \cite{Bulanov:1999} and has been recorded experimentally \cite{Kando:2009}. The radiation released by the solitons propagates essentially in the forward direction, where the seek-after {\it partner photon} signals are expected to be in the EUV range. Thus such soliton-induced radiation signals may not render competing backgrounds to our experiment. Nevertheless, we will further investigate this collective soliton effect to determine whether some of such radiation might be reflected backwards so as to confuse the Hawking photons, whose wavelengths are indeed close.\\

Our next step is to estimate the radiation with the corresponding process according to these categories, and compare the result with the radiation spectrum generated by the PIC simulation code and assess their impacts on the AnaBHEL experiment.

\section{Strategy of AnaBHEL}
We execute the AnaBHEL project based on the following strategy. \\

\noindent
Stage-1

R\&D of the key components, namely the superconducting nanowire single-photon Hawking detector and the supersonic gas jet with designed density profile, are mainly carried out at Leung Center for Cosmology and Particle Astrophysics (LeCosPA), National Taiwan University. These have been going well as reported in the previous sections. \\

\noindent
Stage-2

Dynamics of laser-induced plasma mirror trajectory and its correspondence with the plasma density profile. The first attempt has been scheduled at Kansai Photon Science Institute (KPSI) in Kyoto, Japan, using its PW laser facility, in the summer of 2022. We expect that iterative interplays between the gas jet design and the laser-plasma interaction data acquisition are indispensable. \\ 

\noindent
Stage-3

Full-scale analog black hole experiment to detect Hawking and partner photons will be pursued when the Hawking detector has been fully developed and 
the plasma mirror trajectory has been characterized. It is our desire that the Stage-3 experiment would be carried out at the Apollon Laser Facility in Saclay, France.

\section{Conclusion}
The information loss paradox associated with black hole Hawking evaporation is arguably one of the most challenging issues in fundamental physics, because it touches upon a potential conflict between the two pillars of modern physics, general relativity and quantum field theory. Unfortunately, typical astrophysical stellar-size black holes are too cold and too young to be able to shed light on this paradox. Laboratory investigation of analog black holes may help to shed some light to this critical issue. 

There have been various proposals of analog black holes. Different from the approach of invoking fluids (ordinary and super fluid via Bose-Einstein condensate) that tries to mimic the curved spacetime related to the black hole environment, our approach attempts to create an accelerating boundary condition to a flat spacetime while relying on its nontrivial interplay with the quantum vacuum fluctuations. We believe that these different approaches have their respective pros and cons, and are complementary to each other. Together, a more complete picture of black hole evaporation would hopefully emerge. 

Since its launching in 2018, the AnaBHEL Collaboration has been undergoing encouraging progress albeit the COVID-19 pandemic. While the R\&D has not yet completed, we are confident that the end is in sight.

\section*{Acknowledgment} The Taiwan team is supported by Taiwan's Ministry of Science and Technology (MOST) under Project Number 110-2112-M-002-031, and by Leung Center for Cosmology and Particle Astrophysics (LeCosPA), National Taiwan University. KK is supported by JSPS KAKENHI with Grant Number JP21H01103. SP is further supported by Taiwan's Ministry of Education under the grant MoE/NTU grant number: 111L104013. The authors are grateful to Computer and Information Networking Center, National Taiwan University for the support of high-performance computing facilities.


\begin{thebibliography}{200}

\bibitem{Hawking:1974sw} 
  S.~W.~Hawking,
  Commun.\ Math.\ Phys.\  {\bf 43}, 199 (1975)
  Erratum: [Commun.\ Math.\ Phys.\  {\bf 46}, 206 (1976)].
  
 \bibitem{Hawking:1976}
 S. W. Hawking, Phys. Rev. D {\bf 14}, 2460–2473, (1976).

\bibitem{Susskind:1993}
L. Susskind, L. Thorlacius, and J. Uglum, Phys. Rev. D {\bf 48}, 3743–3761, (1993), 

\bibitem{Almheiri:2012a}
A.~Almheiri, D.~Marolf, J.~Polchinski and J.~Sully,
JHEP \textbf{02} (2013), 062
[arXiv:1207.3123 [hep-th]].

\bibitem{Almheiri:2012b}
Almheiri, M, J. Polchinsky, Stanford... (2012).

\bibitem{Mathur:2009}
S. D. Mathur, Class. Quant. Grav. 26, 224001, (2009), [arXiv:0909.1038].

\bibitem{Chen:2015gux}
P.~Chen, Y.~C.~Ong, D.~N.~Page, M.~Sasaki and D.~Yeom,
Phys. Rev. Lett. \textbf{116} (2016) no.16, 161304
[arXiv:1511.05695 [hep-th]].

\bibitem{Buosso:2017}
R. Bousso and M. Porrati, Class. Quant. Grav. 34, no.20, 204001 (2017) [arXiv:1706.00436 [hep-th]].

\bibitem{Giddings:2019}
S. B. Giddings, Phys. Rev. D 100, no.12, 126001 (2019) [arXiv:1903.06160 [hep-th]]. 

\bibitem{Hawking:2015}
S. Hawking, M. Perry, A. Strominger, Phys. Rev. Lett. (2015).

\bibitem{Chen:2014jwq}
P.~Chen, Y.~C.~Ong and D.~Yeom,
Phys. Rept. \textbf{603} (2015), 1-45
[arXiv:1412.8366 [gr-qc]].

  \bibitem{Almheiri:2019psf}
  A.~Almheiri, N.~Engelhardt, D.~Marolf and H.~Maxfield,
  JHEP \textbf{12}, 063 (2019)
  [arXiv:1905.08762 [hep-th]].

\bibitem{Almheiri:2019hni}
A.~Almheiri, R.~Mahajan, J.~Maldacena and Y.~Zhao,
JHEP \textbf{03}, 149 (2020)
[arXiv:1908.10996 [hep-th]].

\bibitem{Penington:2019kki}
G.~Penington, S.~H.~Shenker, D.~Stanford and Z.~Yang,
[arXiv:1911.11977 [hep-th]].

  \bibitem{Almheiri:2019qdq}
  A.~Almheiri, T.~Hartman, J.~Maldacena, E.~Shaghoulian and A.~Tajdini,
  JHEP \textbf{05}, 013 (2020)
  [arXiv:1911.12333 [hep-th]].

  \bibitem{Chen:2021jzx}
  P.~Chen, M.~Sasaki, D.~Yeom and J.~Yoon,
  `Solving information loss paradox via Euclidean path integral,''
  [arXiv:2111.01005 [hep-th]].
 

\bibitem{Unruh:1981}
W. G. Unruh, Phys. Rev. Lett. {\bf 46}, 1351 (1981); W. G. Unruh, Phys. Rev. D {\bf 51}, 2827 (1995).

\bibitem{Schuthold:2005}
R. Schützhold and W. G. Unruh, Phys. Rev. Lett. {\bf 95}, 031301 (2005).


\bibitem{Yablonovitch:1989}
E. Yablonovitch, Phys. Rev. Lett. 62, 1742 (1989).

\bibitem{Belgiorno:2010}
F. Belgiorno et al., Phys. Rev. Lett. 105, 203901 (2010).

\bibitem{deNova:2019}
M. de Nova, J. R. Golubkov, K. Kolobov, et al., Nature 569, 688 (2019).

\bibitem{Chen-Tajima}
P. Chen and T. Tajima, Phys. Rev. Lett. {\bf 83}, 256 (1999).

\bibitem{Fulling:1976}
S. A. Fulling and P. Davies, Proc. R. Soc. London A348, 393 (1976).

\bibitem{Chen:2017}
P. Chen and G. Mourou. Phy. Rev. Lett. {\textbf 118}, 045001 (2017).

\bibitem{Chen:2020}
P. Chen and G. Mourou, Phys. Plasmas {\textbf 27}, 123106 (2020).

\bibitem{EPR}
A. Aspect, P. Grangier, G. Roger, Phys. Rev. Lett. {\bf 49}, 91 (1982). 

\bibitem{Bulanov:2003}
S. V. Bulanov, T. Esirkepov, T. Tajima, Phys. Rev. Lett. {\bf 91}, 085001 (2003).

\bibitem{Naumova:2004}
N. M. Naumova, et al., Phys. Rev. Lett. {\bf 92}, 063902-1 (2004).

\bibitem{Bulanov:2013}
S. Bulanov, et al., Physics-Uspekhi 56, 429 (2013).

\bibitem{Pirozhkov:2007}
A. Pirozhkov et al., Phys. Plasmas 14, 123106 (2007).

\bibitem{Kando:2007}
M. Kando, Y. Fukuda, A. S. Pirozhkov, J. Ma, I. Daito et al., Phys. Rev. Lett. {\bf 99}, 135001 (2007).

\bibitem{Kando:2009}
M. Kando, A. S. Pirozhkov, K. Kawase, T. Zh. Esirkepov, Y. Fukuda et al., Phys. Rev. Lett. {\bf 103}, 235003 (2009).

\bibitem{Wilczek:1992}
F. Wilczek, ``Quantum Purity at a Small Price: Easing a Black Hole Paradox", Proc. Houston Conf. Black Holes (1992).

\bibitem{Hotta:2015}
M. Hotta, R. Schutzhold, W. G. Unruh, Phys. Rev. D {\textbf 91}, 124060 (2015). 

\bibitem{Bardeen:2014}
J. Bardeen, arXiv:1406.4098 (2014).

\bibitem{Page:1993}
D. N. Page, Phys. Rev. Lett. 71, 3743 (1993) [arXiv:hep-th/9306083].

\bibitem{Hotta-Sugita:2015}
M. Hotta, A. Sugita, Prog. Theor. Exp. Phys., 123B04 (2015) [arXiv:1505.05870 [gr.qc]].

\bibitem{Chen-Yeom:2017}
P. Chen and D.-h. Yeom, Phys. Rev. D {\bf 96}, 025016 (20

\bibitem{Good:2020}
M. R. R. Good, E. V. Linder, Phys. Rev. D {\bf 96}, 215010 (2017); M. R. R. Good, E. V. Linder, Phys. Rev. D {\bf 97}, 065006 (2018).

\bibitem{Tajima:1979}
T. Tajima and J. M. Dawson, Phys. Rev. Lett. (1979).

\bibitem{Chen:1985}
P. Chen, J. M. Dawson, R. Huff, T. Katsouleas, Phys. Rev. Lett. (1985).

\bibitem{Liu:2020}
Y. K. Liu, P. Chen, Y. Fang, “Reflectivity and Spectrum of Relativistic Flying Plasma Mirrors”, Phys. Plasmas. {\bf 10}, 103301 (2021); [arXiv:2012.05769 (2020)].

\bibitem{Davies:1977}
P. C. W. Davies and S. A. Fulling, Radiation from moving mirrors and from black holes, Proc. R. Soc. A \textbf{356}, 237 (1977).

\bibitem{Birrell:1982ix}
N. D. Birrell and P. C. W. Davies, \textit{Quantum Fields in Curved Space}, Cambridge Monographs on Mathematical Physics (Cambridge Univ. Press, Cambridge, UK, 1984).

\bibitem{DeWitt1975}
S. DeWitt, Quantum field theory in curved spacetime, Phys.Rep. \textbf{19}, 295 (1975).

\bibitem{Barton1995}
G. Barton and A. Calogeracos, On the quantum electrodynamics of a dispersive mirror.: I. mass shifts, radiation, and radiative reaction, Ann. Phys. (N.Y.) \textbf{238}, 227 (1995).

\bibitem{Nicolaevici2001}
N. Nicolaevici, Quantum radiation from a partially reflecting moving mirror, Class. Quant. Grav. \textbf{18}, 619 (2001).

\bibitem{Nicolaevici2009}
N. Nicolaevici, Semitransparency effects in the moving
mirror model for Hawking radiation, Phys. Rev. D \textbf{80},
125003 (2009).

\bibitem{Lin2020}
K.-N. Lin, C.-E. Chou, and P. Chen, Particle production by a relativistic semitransparent mirror in (1+3)D Minkowski spacetime, Phys. Rev. D \textbf{103}, 025014 (2021), arXiv:2008.12251 [gr-qc].

\bibitem{Lin2021}
K.-N. Lin and P. Chen, Particle production by a relativistic semitransparent mirror of finite transverse size, arXiv:2107.09033 [gr-qc].

\bibitem{kaganovich:2014}
D. Kaganovich et al., Journal of Applied Physics {\bf 116}, 013304 (2014).

\bibitem{helle:2016}
M. H. Helle et al., Phys. Rev. Lett. {\bf 117}, 165001 (2016).

\bibitem{shcmid:2010}
K. Schmid et al., Phys. Rev. ST Accel. Beams {\bf 13}, 091301 (2010).

\bibitem{FangChiangL:2020}
L. Fang-Chiang et al., Physics of Fluids {\bf 32}, 066108 (2020).

\bibitem{Hsu-hsin Chu:2005}
Hsu-hsin Chu, 2005, “Construction of a 10-TW Laser of High Coherence and Stability and Its Application in Laser-Cluster Interaction and X-Ray Lasers”, PhD thesis, National Taiwan University, Taipei.

\bibitem{golovin:2016}
G. Golovin et al., Nucl. Instrum. Methods Phys. Res., Sect. A {\bf 830}, 375 (2016).

\bibitem{kim:2018}
K. N. Kim et al., J. Korean Phys. Soc. {\bf 73}, 561 (2018).

\bibitem{Fang:2018}
M. Fang et al., Plasma Phys. Controlled Fusion {\bf 60}, 075008 (2018).

\bibitem{Hansen:2016}
A. M. Hansen et al.,Rev. Sci. Instrum. {\bf 89}, 10C103 (2018).

\bibitem{Couperus:2016}
J. P. Couperus et al., Nucl. Instrum. Methods Phys. Res., Sect. A {bf 830}, 504 (2016).

\bibitem{Adelmann:2018}
A. Adelmann et al., Appl. Sci. {\bf 8}, 443 (2018).

\bibitem{Epstein:1974}
A. H. Epstein, MIT Gas Turbine Lab Report 117, 1974.

\bibitem{Hanson:2018}
R. K. Hanson and J. M. Seitzman, {\it Handbook of Flow Visualization} (Routledge, 2018), pp. 225-237.

\bibitem{settles:2001}
G. S. Settles, {\it Schlieren and Shadowgraph Techniques}, (Spinger, Berlin, 2001).

\bibitem{mariani:2020}
R. Mariani et al., Journal of Visualization {\bf 23}, 383-393 (2020).

\bibitem{rhoCentralFoamSolver}
C. J. Greenshields et al.,Int. J. Number. Methods Fluids {\bf 63}, 1 (2010).

\bibitem{Zadeh}
Iman Esmaeil Zadeh et. al.,
Superconducting nanowire single-photon detectors: A perspective on
evolution, state-of-the-art, future developments, and applications 
Appl. Phys. Lett. 118, 190502 (2021).

\bibitem{NbNrec}
B. Korzh et. al.,
Demonstration of sub-3 ps temporal resolution with a
superconducting nanowire single-photon detector, 
Nat. Photonics {\bf 14}, 250–255 (2020). 

\bibitem{WSirec}
B. Korzh et. al.,
Wsi superconducting nanowire single photon detector with a temporal
resolution below 5 ps, 
Conference on Lasers and Electro-Optics 
(Optical Society of America, 2018), p. FW3F.3.

\bibitem{Verma1}
V. B. Verma et. al.,
Single-Photon detection in the mid-infrared up to 10 micron 
wavelength using tungsten silicide superconducting nanowire detectors,
APL Photonics 6, 056101 (2021).

\bibitem{Wollman2}
E. Wollman et. al.,
Recent advances in superconducting nanowire single-photon detector
technology for exoplanet transit spectroscopy in the mid-infrared,
J. of Astronomical Telescopes, Instruments, and Systems, 7(1), 011004 (2021)

\bibitem{Wollman1}
E. Wollman et. al.,
Kilopixel array of superconducting nanowire single-photon detectors,
Opt. Express, {\bf 27} 35279 –35289 (2019).

\bibitem{Stupakov}
G. Stupakov,
Phys. Rev. Accel. Beams {\bf 21}, 041301 (2018).

\bibitem{SSSolution}
J. B. Rosenzweig et al.,
Phys. Rev. Lett. {\bf 61}, 98 (1988).

\bibitem{Classification}
S. Markidis et al.,
Automatic Particle Trajectory Classification in Plasma Simulations,
in 2020 IEEE/ACM Workshop on Machine Learning in High Performance Computing Environments (MLHPC) and Workshop on Artificial Intelligence and Machine Learning for Scientific Applications (AI4S), p. 64-71 (2020).


\bibitem{Smilei}
J. Derouillat et al.,
``SMILEI: A collaborative, open-source, multi-purpose particle-in-cell code for plasma simulation",
Comp. Phys. Comm. {\bf 222}, 351-373 (2018).

\bibitem{scikit-learn}
Pedregosa et al.,
Scikit-learn: Machine Learning in Python,
J. Machine Learning Res., {\bf 12}, 2825 (2011).

\bibitem{Bulanov:1999}
S. V. Bulanov, T. Zh. Esirkepov, N. M. Naumova, F. Pegoraro, and V. A. Vshivkov, Phys. Rev. Lett. {\bf 82}, 3440. (1999).


\bibitem{Kando:2009}
M. Kando et al., Euro. Phys. J. D {\bf 55}, 465 (2009); M. Kando et al., AIP Conf. Proc. {\bf 1153}, 61 (2009).



\end{thebibliography}
\end{document}